\documentstyle[emulate_apj,apjfonts,epsf]{article}

\def \farcs{\hbox{$.\!\!^{\prime\prime}$}}

\def \surfsun{\hbox{${\rm M}_\odot{\rm pc}^{-2}$}}

\lefthead{Hoekstra et al.}
\righthead{Weak lensing analysis of MS~1054-03}
\begin{document}

\slugcomment{Accepted for publication in the ApJ}
\title{Hubble Space Telescope weak lensing study of the $z=0.83$ cluster
MS~1054-03$^1$}

\author{H.~Hoekstra$^\star$, M.~Franx$^\dagger$, K.~Kuijken$^\star$}
\affil{ $^\star$~Kapteyn Astronomical Institute, University of Groningen, \\
        P.O. Box 800, 9700 AV Groningen, The Netherlands \\
        e-mail: hoekstra, kuijken@astro.rug.nl\\
	~~\\
	$^\dagger$~Leiden Observatory, \\
	P.O. Box 9513, 2300 RA Leiden, The Netherlands \\
	e-mail: franx@strw.strw.leidenuniv.nl}

\begin{abstract}

We have measured the weak gravitational lensing of faint, distant background 
galaxies by MS~1054-03, a rich and X-ray luminous cluster of galaxies at a 
redshift of $z=0.83$, using a two-colour mosaic of deep WFPC2 images.
The small corrections for the size of the PSF and the high number 
density of background galaxies obtained in these observations result in
an accurate and well calibrated measurement of the lensing induced distortion.

The strength of the lensing signal depends on the redshift distribution 
of the background galaxies. We used photometric redshift distributions from 
the Northern and Southern Hubble Deep Fields to relate the lensing signal to 
the mass. The predicted variations of the signal as a function of apparent 
source magnitude and colour agrees well with the observed lensing signal. 
The uncertainty in the redshift distribution results in a 10\% systematic 
uncertainty in the mass measurement.

We determine a mass of $(1.2\pm0.2)\times 10^{15}~h_{50}^{-1}~{\rm M}_\odot$ 
within an aperture of radius 1~$h_{50}^{-1}$~Mpc. Under the assumption of an 
isothermal mass distribution, the corresponding velocity dispersion is 
$1311^{+83}_{-89}$. For the  mass-to-light ratio we find $269\pm37 h_{50} 
{\rm M}_\odot / {\rm L}_{B\odot}$ after correcting for pass-band and
luminosity evolution. The errors in the mass and mass-to-light ratio 
include the contribution from the random intrinsic ellipticities 
of the source galaxies, but not the (systematic) error due to the uncertainty 
in the redshift distribution. However, the estimates for the mass and 
mass-to-light ratio of MS~1054-03 agree well with other estimators, 
suggesting that the mass calibration works well. 

The reconstruction of the projected mass surface density shows a complex 
mass distribution, consistent with the light distribution. The results 
indicate that MS~1054-03 is a young system. The timescale for relaxation 
is estimated to be at least 1~Gyr. 

We have also studied the masses of the cluster galaxies, by
averaging the tangential shear around the cluster galaxies.
Using the Faber-Jackson scaling relation, we find the velocity dispersion 
of an $L_*$ galaxy (${\rm L}_B=8\times 10^{10} h_{50}^{-2}{\rm L}_{B\odot}$  
for MS~1054-03) is $203\pm33$ km/s.
\end{abstract}


\section{Introduction}
\footnotetext[1]{Based on observations with the NASA/ESA {\it Hubble 
Space Telescope} obtained at the Space Telescope Science Institute,
which is operated by the Association of Universities for Research
in Astronomy, Inc., under NASA contract NAS 5-26555}

Studying the mass distribution of massive clusters at high redshifts
is important as their existence puts strong constraints on possible 
cosmological models (e.g. Eke et al. 1996; Bahcall \& Fan 1998; Donahue et al.
1998). Furthermore, these systems may be young and thus much can be 
learned about the formation of massive clusters of galaxies.

Massive structures in the universe induce a systematic distortion
of the images of faint background sources, providing a powerful
diagnostic to study their mass distributions (e.g. Mellier 1999). 
The weak lensing signal allows us to map the projected mass 
surface density of the lens (e.g. Kaiser \& Squires 1993), and determine
its mass. Although the weak lensing mass estimate does not require assumptions
about the dynamical state or geometry, it does require a good estimate
for the redshift distribution of the faint sources.

Since Tyson, Wenk, \& Valdes (1990) succesfully measured the
systematic distortion of faint galaxies by a foreground cluster, 
many massive clusters of galaxies have been studied (e.g. Bonnet, Mellier, 
\& Fort 1994; Fahlman et al. 1994; Squires et al. 1996). These studies 
concentrated on massive clusters with redshifts $z\sim0.2 - 0.5$.

Smail et al. (1994) tried to measure the weak lensing signal for the
optically selected, high redshift cluster Cl~1603+43 $(z=0.89)$. Its
observed velocity dispersion is $1226^{+245}_{-154}$ km/s (Postman,
Lubin, \& Oke 1998). Smail et al. (1994) used source galaxies down
to a limiting magnitude of $I=25$, but failed to detect a signal.
Their results are consistent with a lower mass for the cluster.
This is supported by the cluster's  X-ray luminosity of 
$L_x(0.1-2.4~{\rm keV})=1.1\times 10^{44} h_{50}^{-2}$ ergs/s 
(Castander et al. 1994), which is an order of magnitude lower than 
the X-ray luminosity of the brightest, high redshift, X-ray selected clusters 

The X-ray emission of clusters of galaxies provides an efficient means
to select massive clusters of galaxies. Since the work of Smail et al. (1994), 
several high redshift, X-ray luminous clusters have been discovered  from 
Einstein (Gioia \& Luppino 1994) and ROSAT observations (Henry et al. 1997; 
Ebeling et al. 1999; Rosati et al. 1998). 

Luppino \& Kaiser (1997; LK97 hereafter) successfully detected the weak
lensing signal from the X-ray selected, high redshift cluster MS~1054-03
using deep ground based imaging. Clowe et al. (1998) reported on the 
detection of weak lensing signals for the X-ray selected, high redshift 
clusters MS~1137+66 and RXJ1717+67, from deep Keck imaging. However,
these studies suffered from the lack of knowledge of the redshift
distribution of the faint sources.

Lensing studies of high redshift clusters are difficult because 
the lensing signal is low and most of the signal comes from small, faint 
galaxies. This is the regime where HST observations have great advantage over 
ground based imaging, because the background galaxies are much better 
resolved. As a result, higher number densities of sources can be reached,
the images are not crowded, and more importantly the correction for the 
circularization by the PSF is much smaller. Especially for high redshift
clusters, where the redshift of the lens approaches that of the sources, 
the lensing signal depends strongly on the redshift distribution of the 
sources. 

In this paper we present our weak lensing analysis of the cluster MS~1054-03 
using a mosaic of deep WFPC2 images.  It is the second cluster we studied 
using a mosaic of deep HST observations. In Hoekstra et al. (1998, HFKS98 
hereafter) we presented the analysis of Cl~1358+62, a rich cluster of 
galaxies at a redshift of $z=0.33$.

MS~1054-03 $(z=0.83)$ is the most distant cluster in the Einstein Extended 
Medium Sensitivity Survey (Gioia et al. 1990; Gioia \& Luppino 1994). It is 
extremely rich and has a restframe X-ray luminosity\footnote{Throughout this 
paper we will use $h_{50}=H_0/(50~{\rm km/s/Mpc})$, $\Omega_m$=0.3 and 
$\Omega_\Lambda=0$. This gives a scale of $1''=9.5~h_{50}^{-1}$ kpc at the 
distance of MS~1054.} 
$L_x(2-10~{\rm keV})=2.2\times 10^{45} h_{50}^{-2}$ ergs/s (Donahue et al. 
1998), which makes it one of the brightest X-ray clusters known. From their 
ASCA observations Donahue et al (1998) find a high X-ray temperature of 
$12.3^{+3.1}_{-2.2}$ keV.

Our space based observations allow us to study the cluster mass distribution
in much more detail than LK97. Photometric redshift distributions inferred 
from the Hubble Deep Field North (Fern\'{a}ndez-Soto, Lanzetta, \& Yahil 1999) 
and South (Chen et al. 1998) give a good approximation of the redshift 
distribution of the sources used in this weak lensing analysis. As a result
it is possible to obtain well calibrated mass estimates for high redshift
clusters of galaxies from multi-colour imaging, even if the availability of 
spectroscopic redshift data for the sources is limited.

The observations and data reduction are outlined in section~2. In 
section~3 we briefly discuss the method we use to analyse the galaxies. 
Section~4 deals with the light distribution of the cluster. In section~5 we 
present the measurements of the lensing induced distortions of the faint 
galaxies and the reconstruction of the projected mass surface density. 
Section~6 deals with the redshift distribution of the faint galaxies.
Our mass determination is presented in section~7. In this section we
also discuss several effects that complicate accurate mass determinations.
The results for the mass-to-light ratio are given in section~8. In section~9 
we examine in more detail the substructure of the cluster. The lensing signal 
due to the cluster galaxies themselves is studied in section~10.

\section{Observations}

In this analysis we use WFPC2 images taken with the Hubble Space Telescope 
(GO proposal 7372, PI Franx). Figure~\ref{layout} shows the layout of the 
mosaic constructed from the 6 pointings of the telescope. The integration 
time per pointing was 6400s in both the $F606W$ and $F814W$ filter. For the 
weak lensing analysis we omitted the data of the Planetary Camera because of 
the brighter isophotal limit. The total area covered by the observations is 
approximately 26 arcmin$^2$.

\vbox{
\begin{center}
\leavevmode
\hbox{%
\epsfxsize=7.6cm
\epsffile{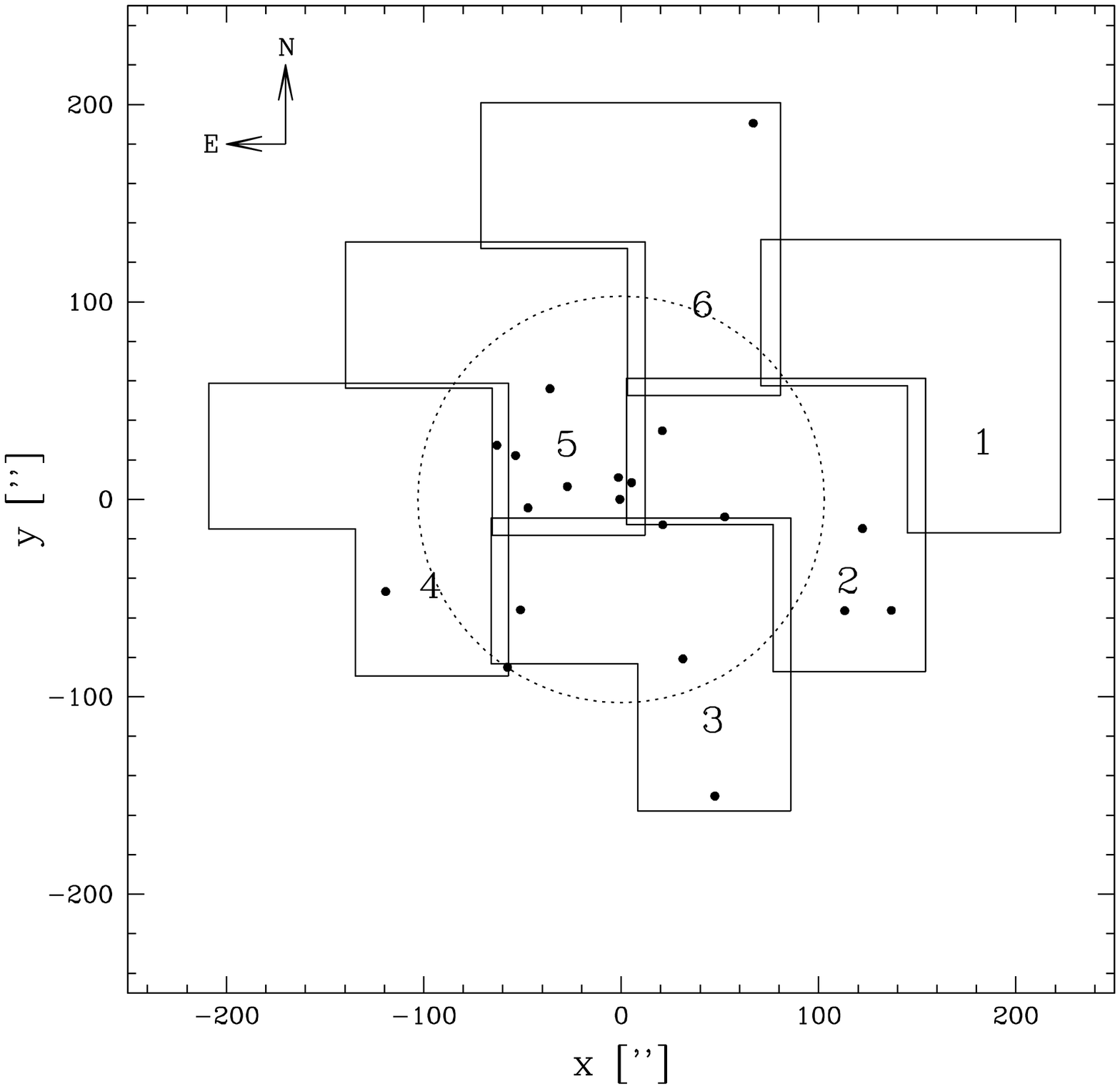}}
\begin{small}
\figcaption{Layout of the observed field. The six pointings are indicated
and numbered. The positions of the 20 brightest cluster galaxies are indicated 
by the black dots. The dashed circle indicates an aperture with a radius of 1 
$h_{50}^{-1}$ Mpc. The image is oriented such that North is up. The area 
covered by the observations is approximately 26 arcmin$^{-2}$.
\label{layout}}
\end{small}
\end{center}}

For each pointing the exposures of a given passband were split into two sets. 
The images in each set were offset from each other by integer pixels 
($\pm 3$ pixels). Both sets were reduced and combined separately, resulting 
in two images (for details see van Dokkum et al. 1999; van Dokkum 1999).
These two images were designed to be offset by 5.5 pixels, to allow the 
construction of an interlaced image (e.g. Fruchter \& Hook 1998; 
van Dokkum 1999), which has a sampling that is a factor $\sqrt{2}$ better than 
the original WFPC2 images. 

Except for the images in the first pointing in F814W, all offsets were within 
0.1 pixel of the requested value. We found that we had to omit the shape 
measurements in the $F814W$ band for pointing~1, because of the deviating 
offset for this exposure. However, we could use the $F606W$ image, and as the 
noise in the shear estimate is dominated by the intrinsic ellipticities 
of the galaxies, the loss in signal-to-noise is minimal.

\section{Object analysis}

Our analysis technique is based on that developed by Kaiser, Squires, 
\& Broadhurst (1995; KSB95 hereafter) and LK97, with a number of 
modifications which are described in HFKS98.

The first step in the analysis is the detection of the galaxy images, for 
which we used the hierarchical peak finding algorithm from KSB95. We selected 
objects with a significance $\nu$ larger than $5\sigma$ over the local sky. 
The detected objects were analysed, yielding estimates for the sizes, 
magnitudes and shapes of the objects. The detection and analysis were done per 
WFPC2 chip and filter. As a new feature, our software computes an estimate 
for the error on the polarization from the data. This error estimate is 
discussed in Appendix~A.

The resulting catalogs were inspected visually in order to remove spurious 
detections, like diffraction spikes, HII regions in resolved galaxies, etc. 
4402 objects were detected in the $F606W$ images, whereas the $F814W$ images
yielded 4290 stars and galaxies. The combined catalog consisted of
5848 stars and galaxies, of which 2844 were detected in both filters.

\vbox{
\begin{center}
\leavevmode
\hbox{%
\epsfxsize=8cm
\epsffile{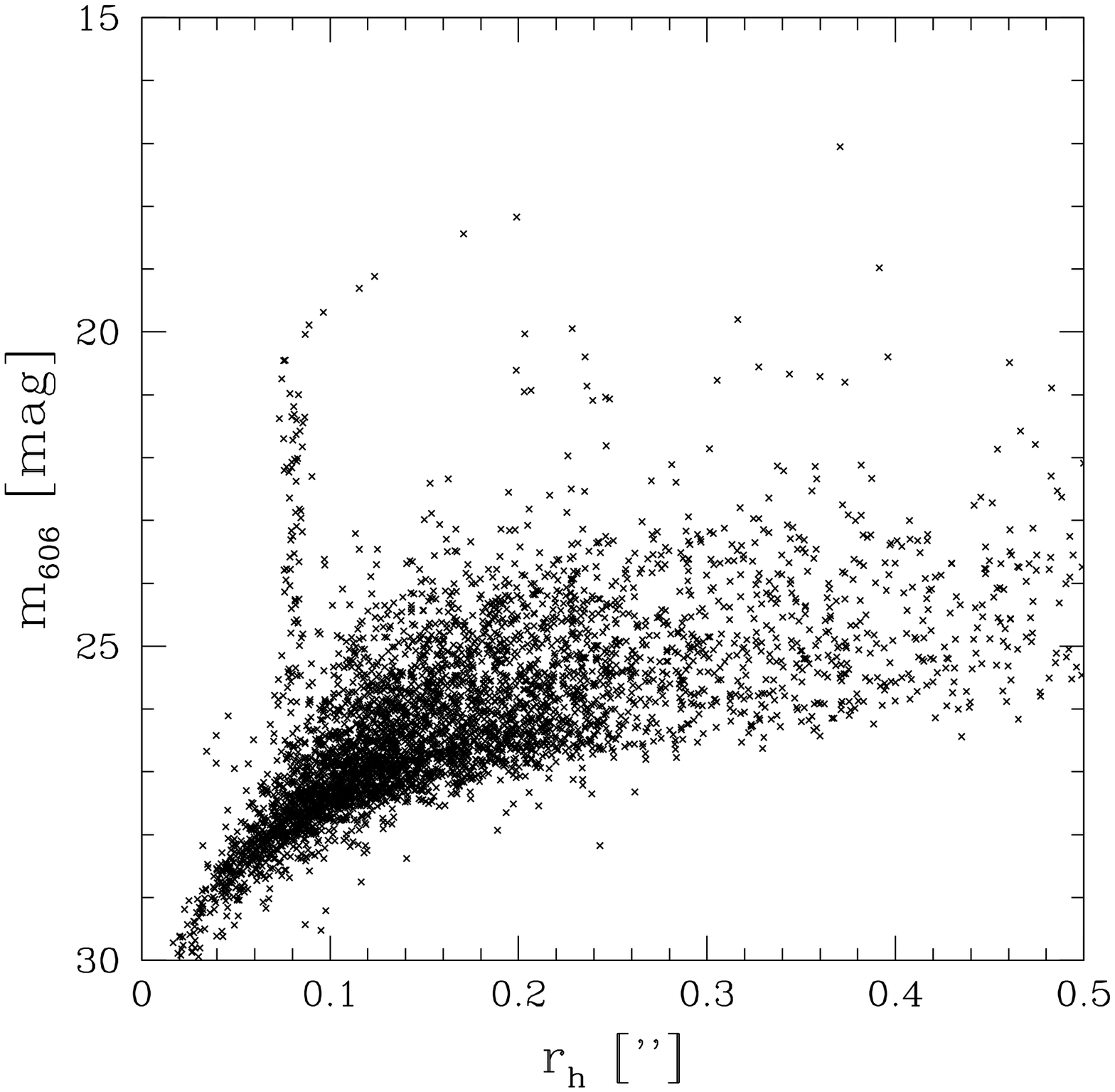}}
\hbox{%
\epsfxsize=8cm
\epsffile{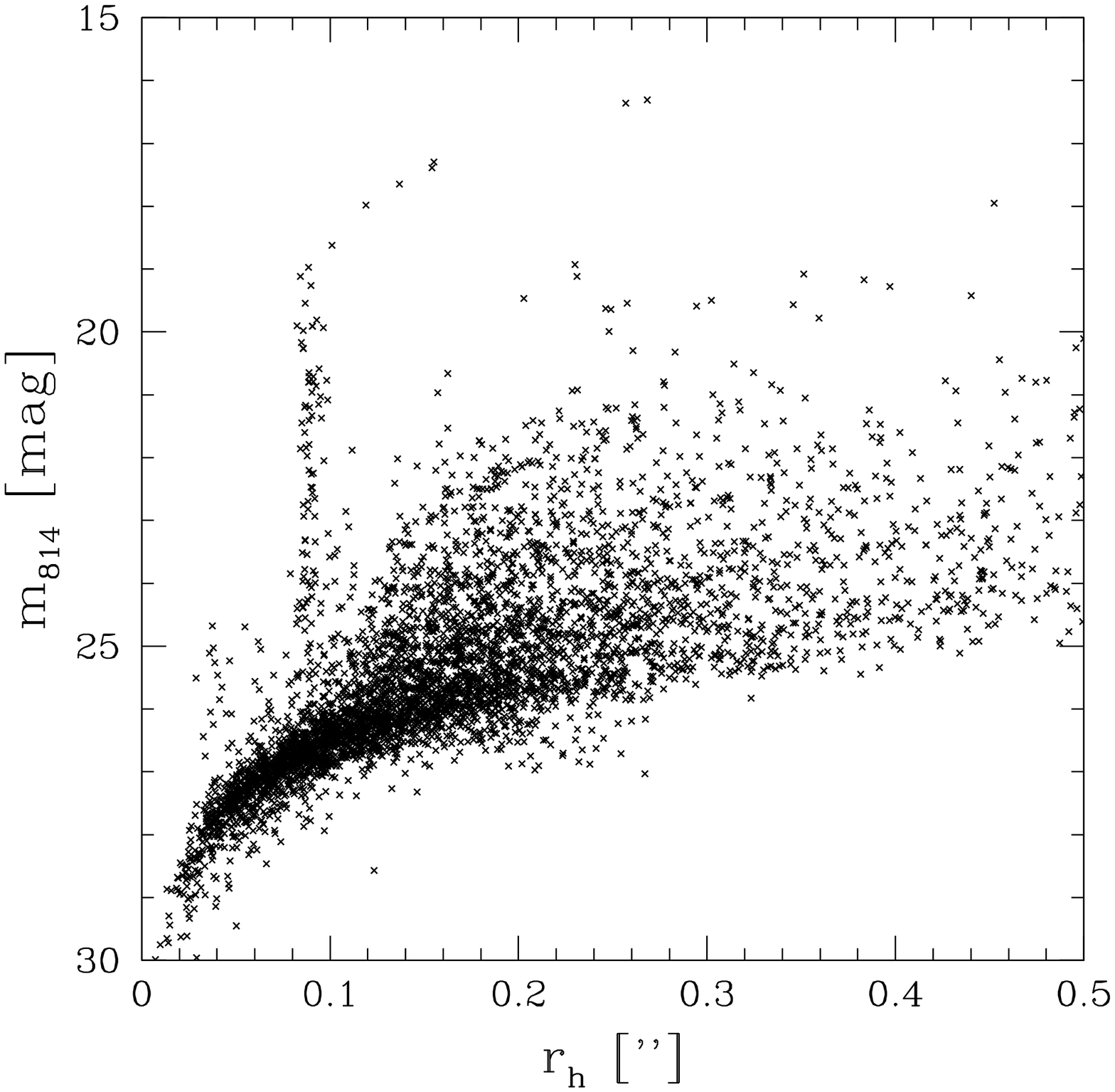}}
\begin{small}
\figcaption{Plot of the apparent magnitude in the F606W filter (upper panel) 
and the F814W filter (lower panel) versus half light radius. In both figures 
the vertical locus where stars are found is easily identified.
\label{objects}}
\end{small}
\end{center}}

For all detected objects we estimated the apparent magnitude using an aperture 
with a diameter of $6\sqrt{2}r_g$ pixels (where $r_g$ is the Gaussian 
scalelength of the object as given by the peak finding algorithm). We 
converted the measured counts to $F606W$ and $F814W$ magnitudes, zero-pointed 
to Vega, using the zeropoints given in the HST Data Handbook (STScI, 
Baltimore). We added 0.05 magnitude to the zero-points to account for the 
Charge Transfer Efficiency (CTE) effect. 

Figure~\ref{objects} shows the apparent magnitude versus the  half 
light radius of the object. Moderately bright stars are easily identified
as they are located in the vertical locus. Bright stars saturate and hence 
their measured sizes increase. Using figure~\ref{objects} we selected a 
sample of moderately bright stars, which were used to examine the PSF.

To obtain a catalog of galaxies we removed the stars from the catalog. At 
faint levels the discrimination between stars and galaxies is not possible, 
but by extrapolating the number counts of moderately bright stars to
faint magnitudes, we find that the stars contribute less than 2\% to the
total counts.

We attempted to determine colours for all detected objects (including
galaxies detected in only one filter), using the same aperture in both 
filters. The software failed to determine colours for some of the galaxies. 
Most of these were extremely faint, and detected only in the $F814W$ band. 
For the weak lensing analysis we only used galaxies for which colours were 
determined. Due to this selection we lost 30\% of the galaxies fainter than 
$F814W=26.5$ (533 galaxies). 

The shape parameters of the galaxies were corrected for the PSF 
(cf. section~3.1) and the camera distortion\footnote{In HFKS98 we 
discussed the correction for the shear introduced by the camera. 
It was subsequently pointed out by Rhodes, Refregier, \& Groth (1999) 
that the camera shear in fact is radial, and not tangential as presented 
in HFKS98. Because the effect is small, and we used a mosaic of images, 
it does not change the results of HFKS98.} We used the estimated errors on the 
shape measurements (cf. Appendix~A) to combine the results from the
$F606W$ and $F814W$ images in an optimal way, resulting in a final catalog 
of 4971 galaxies. 

The better sampling of the interlaced images improved the object
detection and analysis significantly. HFKS98 showed that due to the
poor sampling of WFPC2 images only the shapes of galaxies with sizes
$r_g>0\farcs 12$ could be measured reliably. The images used here
allow us to use all detected galaxies. HFKS98 found a number density of 
79 galaxies arcmin$^{-2}$ from 3600s exposures in both $F606W$ and $F814W$. 
The observations presented here yield a number density of 191 galaxies 
arcmin$^{-2}$. This higher number density is due to both the longer 
integration time, and the improved sampling of the interlaced images.
An analysis of the effect of small pointing errors in the interlaced
images shows that these are not important for our analysis (cf. Appendix~B).

Figure~\ref{counts} shows the observed number counts of galaxies in
our observations versus apparent magnitude. For comparison we also 
show the counts from the HDF-North (Williams et al. 1996). This indicates 
that our catalog of galaxies is complete down to $F606W=26.5$ and $F814W=26$.

\subsection{Corrections}

Several observational effects systematically distort the images of the
galaxies. To obtain an accurate and unbiased estimate of the weak lensing 
signal these effects have to be corrected for. To do so, we follow the
scheme outlined in KSB95, LK97, and HFKS98. 

\vbox{
\begin{center}
\leavevmode
\hbox{%
\epsfxsize=8cm
\epsffile{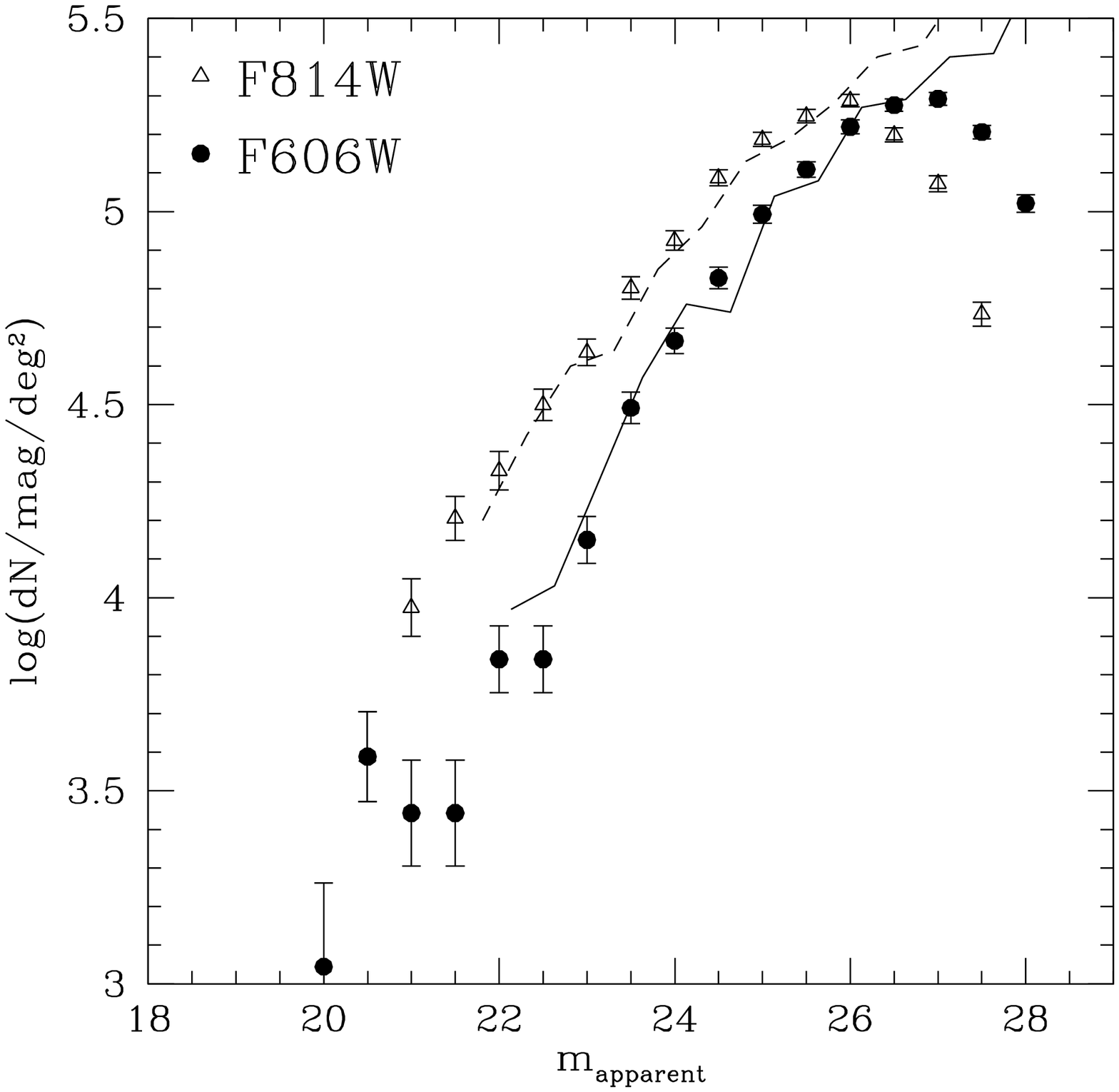}}
\begin{small}
\figcaption{Plot of the number counts of detected galaxies in the $F606W$ band 
(solid circles) and $F814W$ band (open triangles). The dashed line shows
the number counts from the HDF-N (Williams et al. 1996) in the $F814W$ filter,
and the solid line shows the counts in the $F606W$ filter. This indicates 
that our catalog is complete down to $F606W=26.5$ and $F814W=26$.
\label{counts}}
\end{small}
\end{center}}

\vspace{0.3cm}

\vbox{
\begin{center}
\leavevmode
\hbox{%
\epsfxsize=8cm
\epsffile{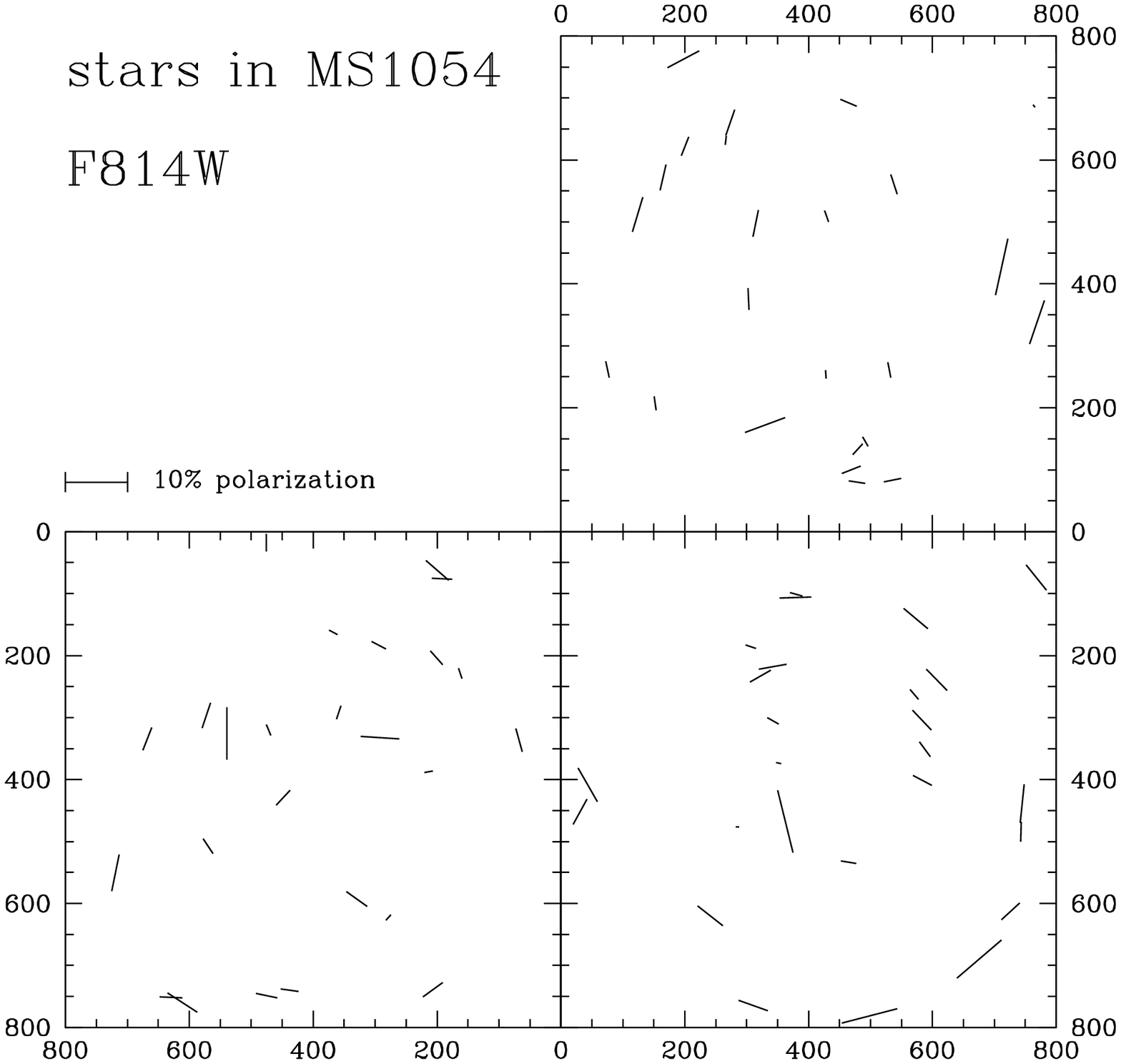}}
\begin{small}
\figcaption{The polarization field of stars taken from the interlaced 
observations in the F814W filter. The orientation of the sticks indicates the 
direction of the major axis of the PSF, whereas the length is proportional to 
the size of the anisotropy. The polarization values are measured using a 
Gaussian weight function with a dispersion of $0\farcs07$. The lower left 
panel corresponds to chip 2, lower right to chip 3, and the upper right one 
denotes chip 4. We have omitted chip 1, which is the planetary camera.
\label{psf814}}
\end{small}
\end{center}}

HFKS98 showed that the PSF of WFPC2 is highly anisotropic at the edges
of the chips. We selected a sample of moderately bright stars from the
observations of MS~1054. Figure~\ref{psf814} shows the observed polarization 
of these stars in the F814W band. The anisotropy pattern is similar to the 
pattern presented in HFKS98.

Because the PSF changes slightly with time, we fitted the shape parameters of 
our new observations to a scaled model of the PSF polarization measured from 
the globular cluster~M4 and added a first order polynomial:
$$ p_i^{\rm new}= a \cdot p_i^{\rm old} + c_0 + c_1 x + c_2 y.$$
This procedure gave excellent fits to the observed PSF anisotropy in our 
observations of MS~1054. 

The PSF also circularizes the images of the small, faint galaxies, thus 
lowering the amplitude of the lensing signal. Compared to ground based 
observations, the small faint galaxies are much better resolved in HST 
observations, resulting in much smaller corrections for the circularization 
by the PSF. This allows us to measure a well calibrated lensing signal.

To obtain estimates for the real shapes of the faint galaxies, we
determined the 'pre-seeing' shear polarizability $P^\gamma$ (LK97; HFKS98). 
The measurements of $P^\gamma$ for individual galaxies are rather noisy,
and therefore we binned the data as a function of size of the object.

\vbox{
\begin{center}
\leavevmode
\hbox{%
\epsfxsize=8cm
\epsffile{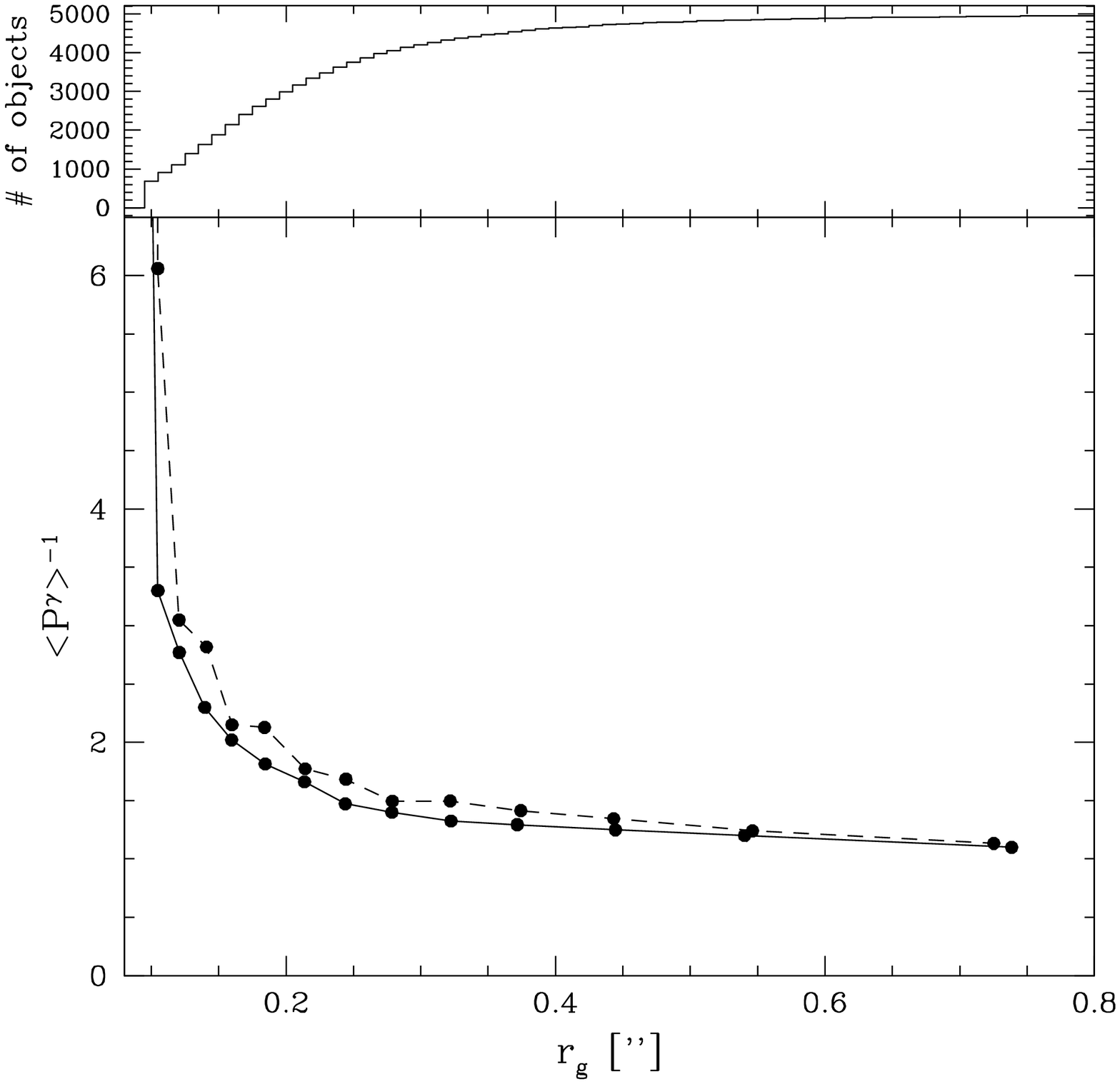}}
\begin{small}
\figcaption{Required correction factor $1/P^\gamma$ to estimate the 
'pre-seeing' ellipticity of the object. The solid line shows the results for 
the $F606W$ filter, and the dashed line shows the correction factor for 
$F814W$. The pre-seeing shear polarizability $P^\gamma$ corrects for the 
circularization by the weight function {\it and} the PSF. It depends strongly 
on the size of the object, and increases rapidly as the object's size 
approaches the size of the PSF. As the PSF in ground based data is much larger,
the corrections for galaxies of similar sizes would be much larger.
The upper panel shows the cumulative number of objects as a function of size.
The objects for which the peak finder assigned a size smaller than the PSF
are analysed with a fixed weight function matched to the PSF. These
objects end up in the bin of the smallest objects.
\label{corfac}}
\end{small}
\end{center}}

\begin{figure*}[b!]
\begin{center}
\leavevmode
\hbox{%
\epsfxsize=\hsize
\epsffile[18 144 592 448]{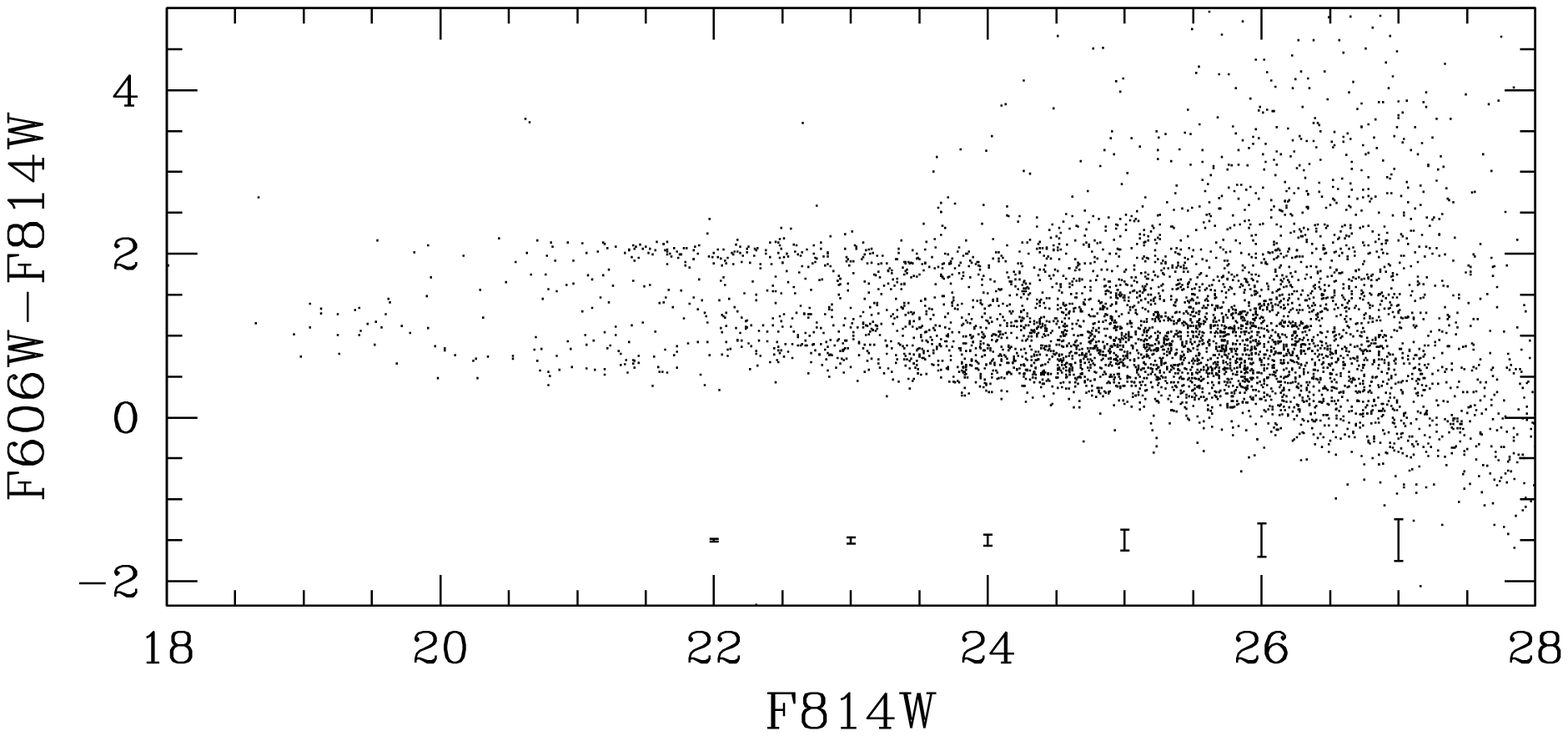}}
\begin{small}
\figcaption{Colour-magnitude diagram of the objects for which colours could
be determined. The cluster colour-magnitude relation is clearly 
visible. The errorbars at the bottom of the plot correspond to the 
errorbars on the $F814W$ magnitude.
\label{colmag}}
\end{small}
\end{center}  
\end{figure*}

The results are presented in figure~\ref{corfac}, which shows the 
required correction factor to obtain the true ellipticity of a galaxy from 
the observed polarization. The upper panel shows the cumulative number of 
objects as a function of object size. The objects for which the peak finder 
assigned a size smaller than the PSF were analysed with a weight function 
matched to the PSF. These objects end up in the bin corresponding to the 
smallest objects.

The correction for objects with sizes comparable to the size of the PSF
is large. The PSF in ground based images is much larger, and therefore
large correction factors are required for objects that are still well 
resolved in HST images. From a $0\farcs65$ deep ground based image
we found that $P^\gamma\approx 0.2$ for galaxies of $R\sim25$, whereas
galaxies of similar brightness in our HST images have $P^\gamma\approx 0.6$.

As a test of our procedure we compared the distortion measured from 
the $F606W$ images and the $F814W$ images (which included the correction 
for the circularization) using only objects for which shape parameters
could be measured in both images. We compared the average tangential
distortions and found that the results for the two filters agree well to
one another (we found $\langle g_T \rangle=0.0366\pm0.0066$ from the $F606W$ 
data and $\langle g_T\rangle=0.0375\pm0.0068$ for the $F814W$ images).

Recently it has been found that the Charge Transfer Efficiency problem for
WFPC2 has worsened. This can systematically change the shapes of the faint
galaxies, as charge is 'smeared' along the CCD columns. This could result
in a detectable change in the shapes of the galaxy images.

We compared the shapes of the objects in the regions where the pointings
overlap (cf. figure~\ref{layout}). Unfortunately the number of objects was 
low, and the contribution of shot noise to the shape estimates substantial.
We did not detect a systematic change in the shapes, but we cannot put strong 
constraints on the size of the effect. We also simulated the effect, assuming 
that the change in shape increased linearly with row number for each chip, 
with no change at low row numbers and a shear of 10\% at high row numbers. 
We found that this resulted in a neglegible change in our mass estimate. 

\section{Light distribution}

Figure~\ref{colmag} shows a plot of the colour of the galaxies versus 
their $F814W$ magnitude. The cluster is visible as the concentration of 
galaxies with $F606W-F814W\sim 2$. The galaxies on this colour-magnitude 
relation are generally early type galaxies, which are among the reddest 
objects in the cluster. 

We selected 324 galaxies brighter than $F814W=24.5$, lying close to the 
colour-magnitude relation, to study the cluster light distribution.
This catalog was compared to the spectroscopic catalog (van Dokkum 1999).
The contamination by field galaxies is small, and we removed galaxies
with redshifts incompatible to that of the cluster. Many of these belong
to a small group at $z=0.547$ (van Dokkum 1999).

\begin{figure*}
\begin{center}
\leavevmode
\hbox{%
\epsfxsize=\hsize
\epsffile{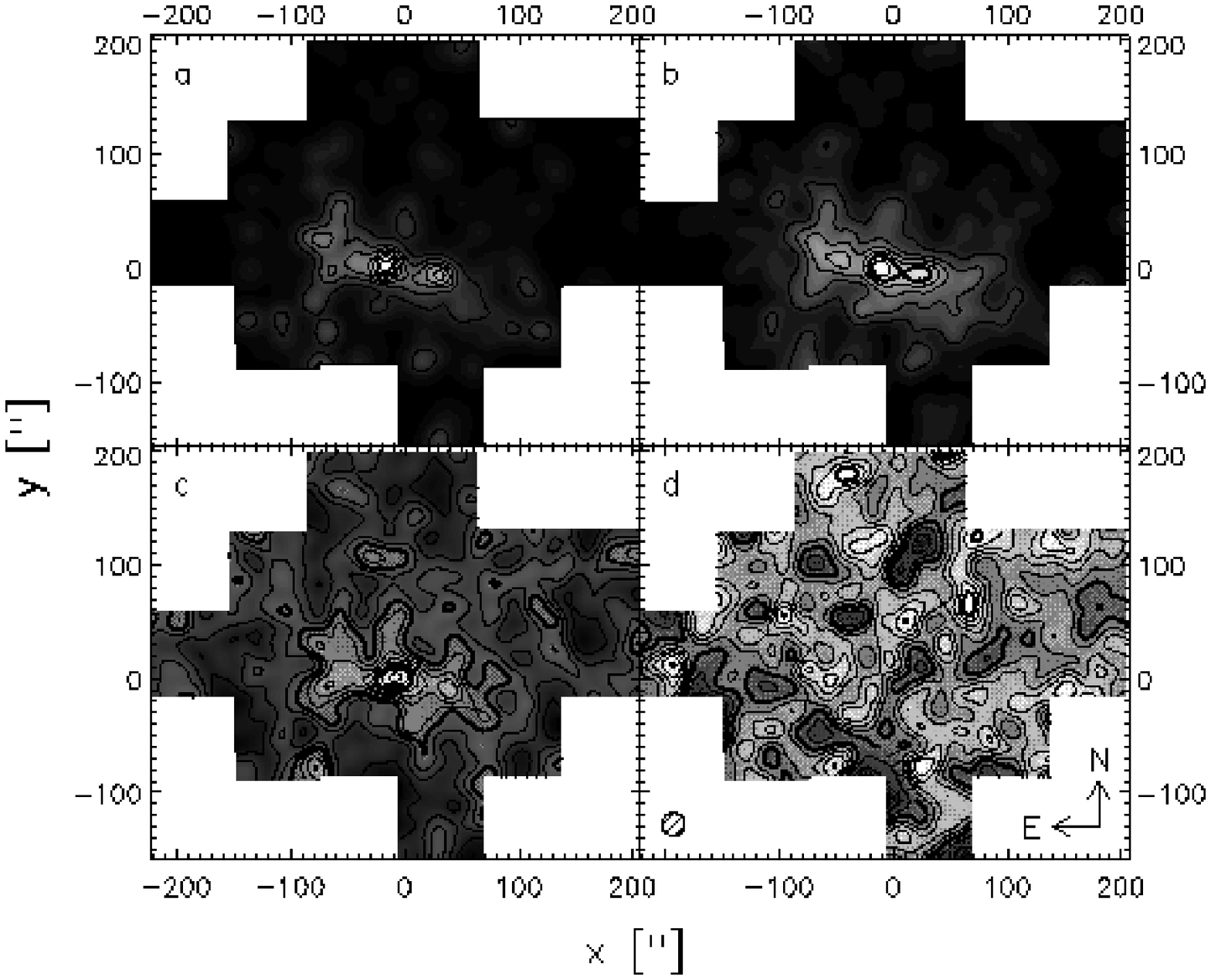}}
\begin{small}
\figcaption{(a) Luminosity distribution of cluster galaxies brighter
than $F814W$ 24.5. The contours are 10\%, 20\%, etc. of the
peak value; (b) Number density distribution of this sample of 
cluster galaxies; (c) Number density distribution of all detected galaxies
brighter than $F814W$ 25; (d) Number density distribution of all detected 
galaxies fainter than $F814W$ 25, where we corrected the counts for
the area covered by the (large) bright galaxies. In panels b--d the interval 
between subsequent contours are 20 galaxies arcmin$^{-2}$, and the thick lines
indicate levels of 100 and 200 galaxies arcmin$^{-2}$. All distributions 
have been smoothed with a Gaussian filter with a FWHM of 20 arcseconds
(indicated by the shaded circle). The mosaic covers approximately  
26 arcmin$^{-2}$, and the total field shown is 7 by 6 arcmin.
\label{lumdis}}
\end{small}
\end{center}  
\end{figure*}

We have use SExtractor (Bertin \& Arnouts 1996) to determine total
magnitudes for these galaxies. Figure~\ref{lumdis} shows grey scale 
plots of the luminosity and number density weighted light distributions. 
The images have been smoothed with a Gaussian with a FWHM of 20 arcseconds. 
The galaxy number density and the luminosity weighted distributions look 
quite similar. The cluster light distribution is elongated east-west, and 
three clumps can be identified.

Figure~\ref{lumdis} also shows gray scale images of the smoothed number
density when we split the sample of all detected galaxies into a bright
($F814W<25$) and a faint sample ($F814W>25$). The cluster is clearly visible 
in the bright sample, but does not show up in the number counts of the faint 
galaxies (which have been corrected for the area covered by large bright
galaxies). Figure~\ref{lumdis}d shows several significant peaks, indicative 
of some clustering at high redshift.

We estimate the luminosity in the redshifted $B$ band following
van Dokkum (1999). The direct transformation from the HST filters to the 
passband corrected $B$ band is given by 

$$ B_z = F814W - 0.03 ( F606W - F814W ) + 1.23,$$

\noindent where $B_z$ denotes the passband corrected $B$ band magnitude. The
luminosity is given by

$$ L_B=10^{0.4(M_{B\odot}-B_z+DM+A_{F814W})} L_{B\odot},$$

where $M_{B\odot}=5.48$ is the solar absolute $B$ magnitude, $DM$ is the
distance modulus, and $A_{F814W}$ is the extinction correction in the
$F814W$ filter towards MS~1054-03. The redshift of $0.83$ for MS~1054
gives a distance modulus of $44.09-5\log h_{50}$. Taking galactic
extinctions from Burstein \& Heiles (1982) we use a value of 
0.03 for $A_{F814W}$.

At faint magnitudes, the colour-magnitude relation is less clearly defined.
To estimate the total light of the cluster, an estimate of the amount of 
light contributed by galaxies fainter than $F814W=24.5$ is required. 
To do so, we fitted a luminosity function to the sample of bright, colour
selected cluster galaxies. We found that a Schechter function with 
$\alpha=-1.0$ and ${\rm M}_B^*=-21.8+5\log h_{50}$ $({\rm L}_*=8\times 10^{10} 
h_{50}^{-2}{\rm L}_{B\odot})$ could fit the observed counts of the sample of 
bright, colour selected cluster galaxies. For nearby clusters one typically 
finds ${\rm M}_B^*=-21.0+5\log h_{50}$ (eg. Binggeli, Sandage, \& 
Tammann 1988), indicating that the galaxies in MS~1054-03 are approximately 
twice as bright as nearby cluster galaxies. This is in good agreement with 
the results of van Dokkum et al. (1998) who studied the fundamental plane in 
MS~1054-03. The luminosity function indicates that we miss approximately 5\% 
of the light when not including the faint galaxies.

MS~1054-03 is a high redshift cluster and one might expect
a relatively high fraction of blue cluster galaxies, which are too blue to
be included in our colour selected sample of cluster galaxies. Comparison of 
our sample to a sample of bright, spectroscopically confirmed members 
(van Dokkum 1999) indicates that we miss 16\% of the total light in the $B$
band.

The cumulative light profile as a function of radius from the centre for the 
sample of bright galaxies is presented in figure~\ref{lumprof}. The profile is 
multiplied by a factor $1.05\times1.16$ to account for the light from 
the faint cluster galaxies and the bluest galaxies. The solid line corresponds
to the observed profile, whereas the dashed line shows a isothermal profile 
for comparison. We estimate the total luminosity within an aperture of 
1~$h_{50}^{-1}$~Mpc to be $1.0\times10^{13} h_{50}^{-2} L_{B\odot}$.

\section{Weak lensing signal}

Due to the intrinsic ellipticities of the galaxies, their individual shape 
estimates are very noisy measurements of the lensing induced distortion.
We therefore need to average the measurements of a large number of galaxies
to obtain a useful estimate for the distortion $g$.

To estimate the average distortion, we weight each object with the inverse 
square of the uncertainty in the distortion, which includes the
contribution of the intrinsic ellipticies of the galaxies and
the shot noise in the shape measurement (cf. Appendix~A):

\vbox{
\begin{center}
\leavevmode
\hbox{%
\epsfxsize=7cm
\epsffile{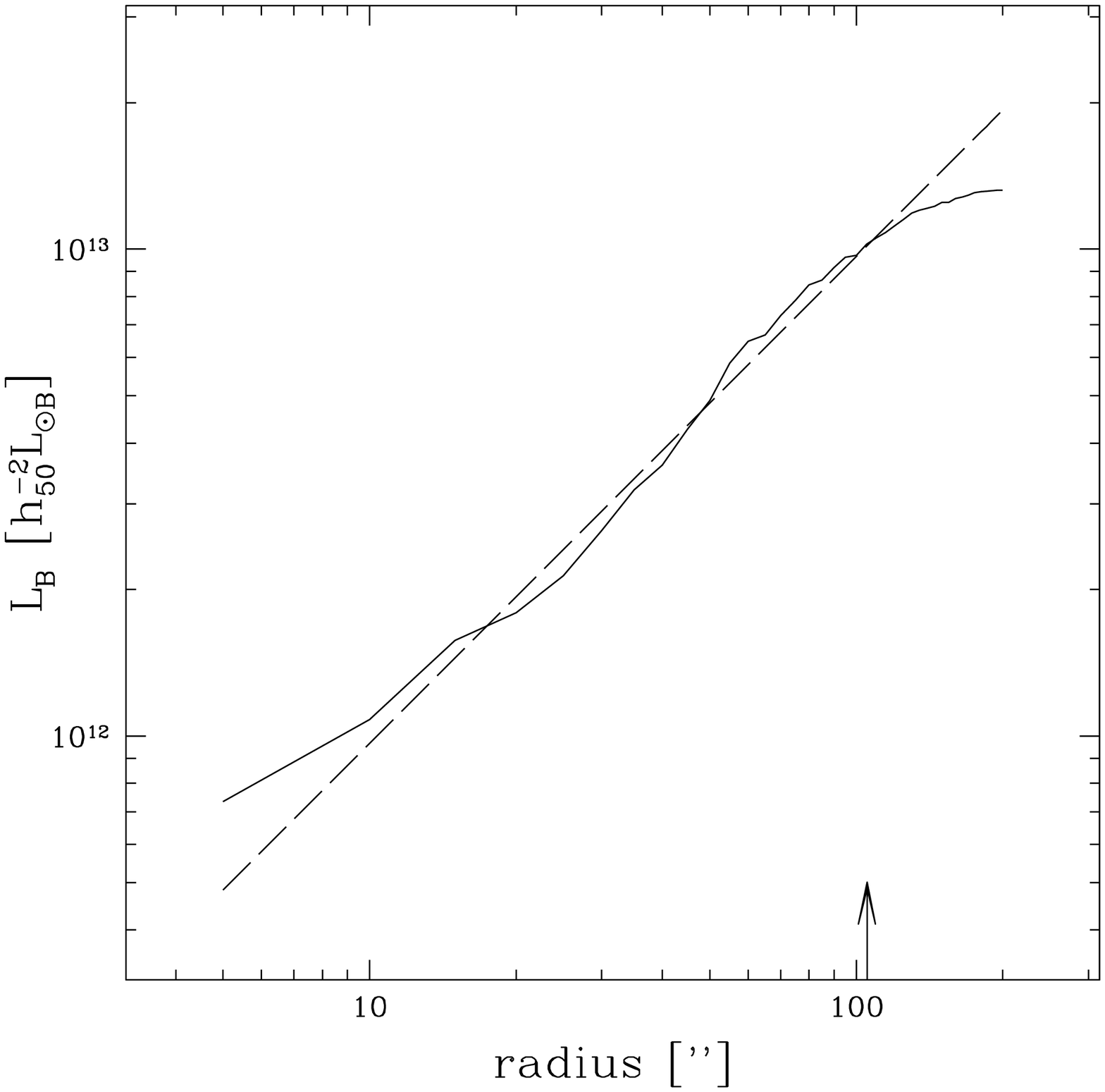}}
\begin{small}
\figcaption{The cumulative, passband corrected, $B$ band luminosity as a 
function of radius from the assumed cluster centre. The solid line
corresponds to the observed profile. The profile was calculated
using the colour selected sample of cluster galaxies, and was 
corrected for the contribution of very blue galaxies and 
faint galaxies. The dashed line corresponds
to an isothermal profile, which is shown for comparison. The
arrow indicates a radius of 1 $h_{50}^{-1}$ Mpc.
\label{lumprof}}
\end{small}
\end{center}}

\begin{equation}
\langle g_i \rangle = \frac{\sum g_{i,n} \sigma_{g,n}^{-2}}
{\sum \sigma_{g,n}^{-2}}.
\end{equation}

Selecting galaxies with $22<F814W<26.5$ and $F606W-F814W<1.6$ 
(the 'source' sample from table~\ref{table1}), and smoothing using a 
Gaussian with a FWHM of 20 arcsec, we obtain the distortion
field presented in figure~\ref{shearfield}. The sticks indicate the 
direction of the distortion, and the length is proportional to the amplitude 
of the distortion. It shows that on average the faint galaxies tend to 
align tangentially to the (elongated) cluster mass distribution, as
expected from gravitational lensing.

The average tangential distortion $g_T$ in radial bins is a useful measure 
of the lensing signal. It is defined as $g_T=-(g_1 \cos 2\phi+g_2\sin 2 \phi)$,
where $\phi$ is the azimuthal angle with respect to the assumed centre of the 
mass distribution. We assume that the position of the Brightest Cluster
Galaxy defines the centre of the cluster. At large radii from the cluster 
centre, where we do not have data on complete circles, we account for possbile
azimuthal variation of $g_T$ by fitting

\begin{equation}
g_T(r,\phi)=\langle g_T \rangle(r) (1+g_{2,c}\cos(2\phi) + 
g_{2,s}\sin(2\phi))\label{gtang}
\end{equation}

\noindent to the observed distortions. This procedure gives the correct value
for an ellipsoidal mass distribution. Although the cluster appears elongated, 
no large azimuthal variations in the tangential distortion were found.

\vbox{
\begin{center}
\leavevmode
\hbox{%
\epsfxsize=8cm
\epsffile[85 226 575 700]{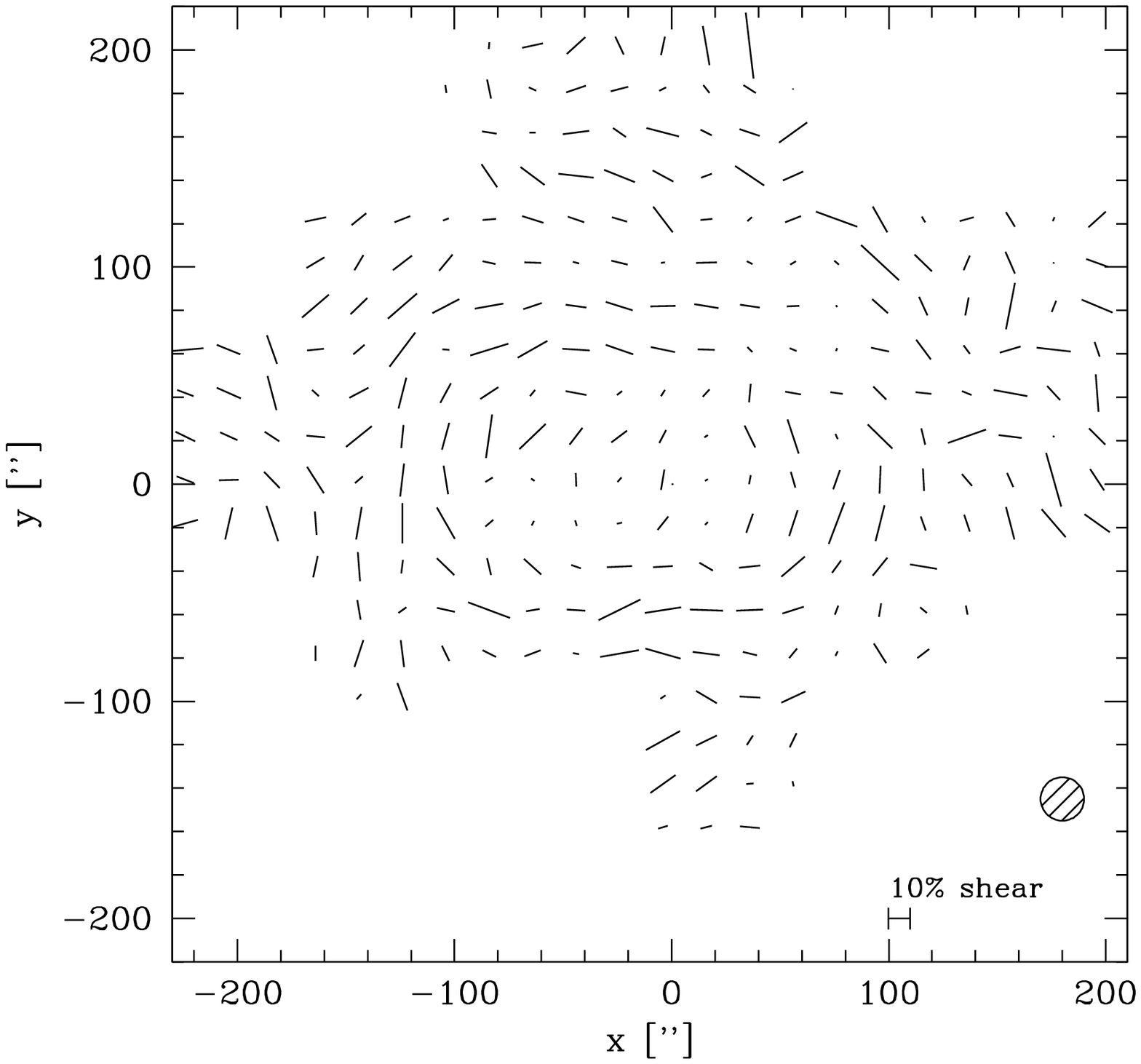}}
\begin{small}
\figcaption{Distortion field $g$ obtained from the sample of
source galaxies, smoothing the data using a Gaussian with a
FWHM of 20 arcsec (indicated by the shaded circle). The orientation 
of the sticks indicates the direction of the distortion, whereas the 
length is proportional to the amplitude of the signal.
Due to gravitational lensing the images of the faint 
sources tend to align tangentially to the cluster mass distribution.
\label{shearfield}}
\end{small}
\end{center}}

\begin{figure*}[b!]
\begin{center}
\leavevmode
\hbox{%
\epsfxsize=\hsize
\epsffile[0 420 600 720]{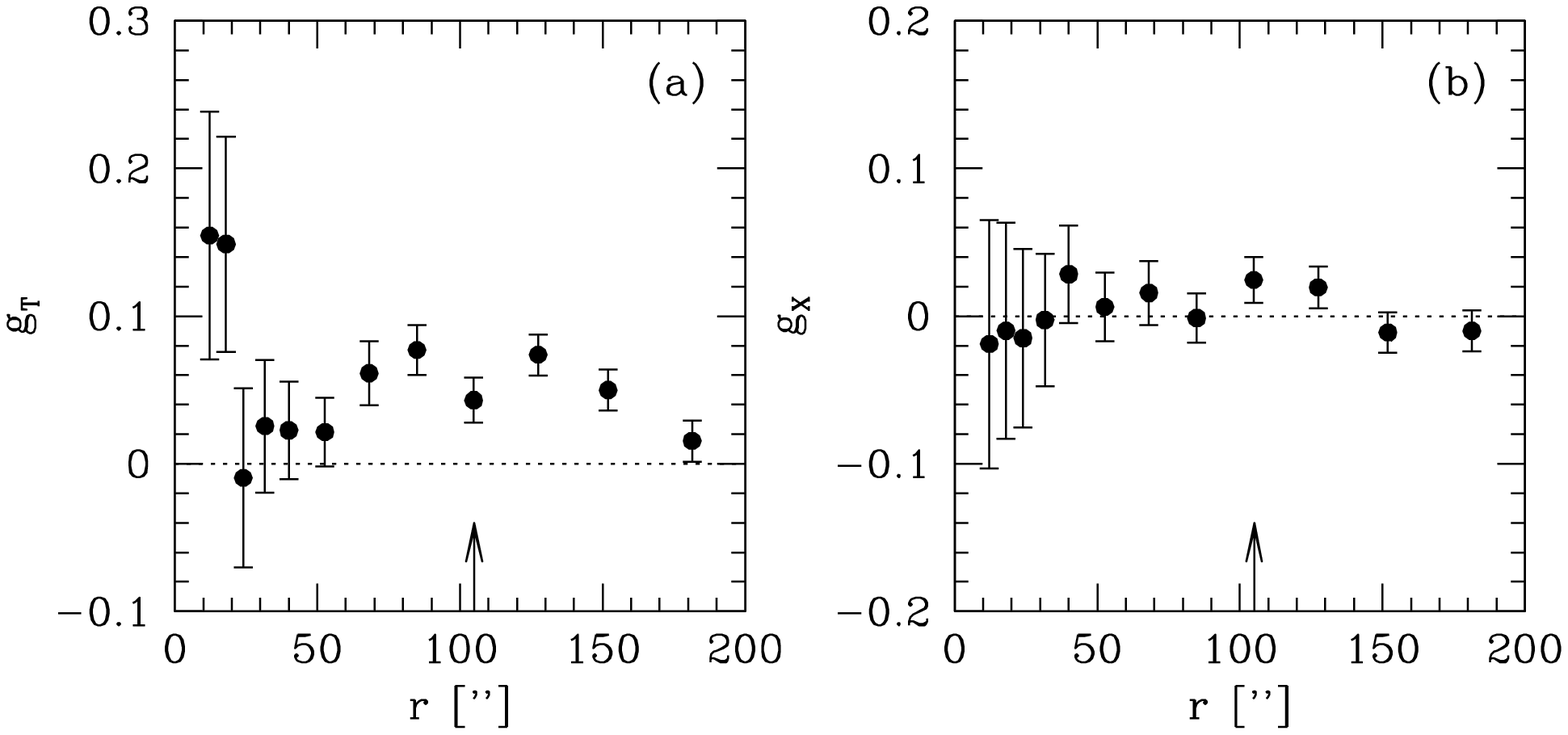}}
\begin{small}
\figcaption{(a) Tangential distortion as a function of radius from 
the cluster centre using the sample of source galaxies as defined
in table~\ref{table1}. A clear lensing signal is detected, although 
the dependence of the signal on radius indicates a complex mass 
distribution. (b) Measured signal when the phase of the distortion is 
increased by $\pi/2$. If the signal observed in (a) is due to gravitational 
lensing, $g_X$ should vanish as is observed. In both figures the arrows 
indicate a radius of $1~h_{50}^{-1}$ Mpc.
\label{gtall}}
\end{small}
\end{center}  
\end{figure*}

The average tangential distortion as a function of radius is presented in 
figure~\ref{gtall}a. In the absence of lensing, the points should scatter 
around zero. The figure shows a systematic tangential alignment of 
the sources. The dependence of the signal on radius is quite complicated, 
indicative of a complex mass distribution. The results of the two-dimensional 
mass reconstruction~(cf. section~5.1) show that the sharp decrease in 
the tangential distortion around 40'' is due to substructure
in the cluster mass distribution.

As a test, figure~\ref{gtall}b shows the results when the phase of the 
distortion is increased by $\pi/2$ (i.e. rotating the sources by $\pi/4$). 
In this case the signal should vanish if it is due gravitational lensing, as
is observed.

We divide our sample of galaxies in several subsamples, based on their
apparent magnitude and colours. Table~\ref{table1} lists some details of 
the subsamples. It shows that even our 'bright' sample in fact consists 
of rather faint galaxies.

\begin{table*}[b!]
\begin{center}
\begin{tabular}{lccccc}
\hline
\hline
(1)		&   (2)	    & (3)	     & (4)         &  (5)     & (6) \\
name            & $F814W$   & $F606W-F814W$  & \# galaxies & $\bar n$ & $\bar m$\\
		& [mag]	    & [mag]	     &             & arcmin$^{-2}$ & [mag]\\
\hline
source		& $22-26.5$ & $< 1.6$		& 2643 & 102 & 25.2 \\
bright          & $22-25$   & $< 1.6$		& 1137 & 44  & 24.2 \\
faint           & $25-26.5$ & $ - $             & 1906 & 73  & 25.8 \\
blue            & $22-26.5$ & $<0.8$            & 1198 & 46  & 25.4 \\
red             & $22-26.5$ & $0.8 - 1.6$   	& 1445 & 55  & 25.1 \\

\hline
\hline
\end{tabular}
\begin{small}
\caption{Properties of the various subsamples taken from
the final catalog of galaxies.\label{table1}}
\end{small}
\end{center}
\end{table*}

The tangential distortion profiles of the various subsamples are presented
in figure~\ref{profiles}. A clear lensing signal is detected for all 
subsamples. The signals of the bright and the faint sample are 
independent measurements, and agree well with one-another. Although the 
profiles of the blue and the red sample agree well, the difference in the 
amplitude of the lensing signal is clear. This is due to the fact that, on
average, the redder galaxies are at lower redshifts than the blue galaxies. 
In section~6 the issue of the redshift distribution of the background galaxies
is discussed in detail.

\vbox{
\begin{center}
\leavevmode
\hbox{%
\epsfxsize=8cm
\epsffile{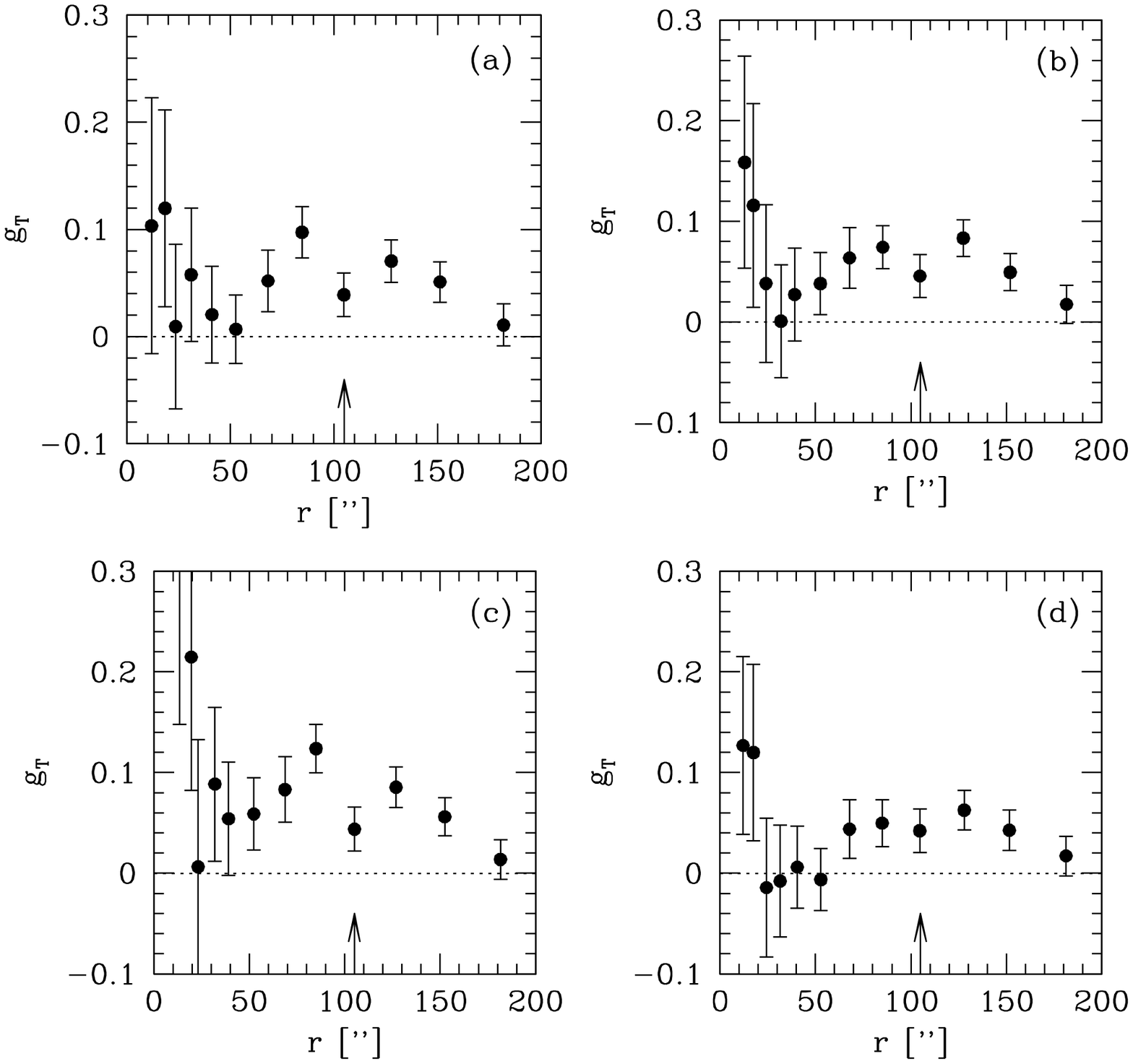}}
\begin{small}
\figcaption{Tangential distortion as a function of radius from the cluster
centre for the various subsamples: (a) bright galaxies; (b) faint galaxies; 
(c) blue galaxies; (d) red galaxies. The signals for the bright and the faint
galaxies are quite similar in both strength and shape. Although the shapes
of the signals for the blue and the red sample are similar, the difference
in strength of the signal is clear. In these figures the arrows indicates a 
radius of $1~h_{50}^{-1}$ Mpc.
\label{profiles}}
\end{small}
\end{center}}

\subsection{Projected mass distribution}

Figure~\ref{massrec} shows the reconstructed mass surface density based
on the distortion field presented in figure~\ref{shearfield}. We used
the maximum probability extension of the original KS algorithm 
(Kaiser \& Squires 1993; Squires \& Kaiser 1996), which does not suffer 
from some of the biases of the original KS93 inversion, and can be applied 
directly to fields with complicated boundaries. Several effects
complicate a direct interpretation of the mass map (cf. section~7.1).
However, if the distortions are relatively small, it gives a fair description 
of the cluster mass distribution.

The mass reconstruction shows that the cluster mass distribution
is quite complex: three distinct peaks are visible. The positions of the 
peaks coincide fairly well with the peaks in the light distribution 
(cf. Fig.~\ref{lumdis}). The results presented in LK97 lacked the resolution
to detect the substructure. 

Donahue et al. (1998) studied the X-ray properties of MS~1054 and found that 
the distribution of the gas is elongated similar to the weak lensing mass 
distribution. They also find evidence for substructure. However, the peak in 
their map of the X-ray brightness coincides with the western clump. Deeper 
X-ray observations by Neumann et al. (1999) show that the peak in the X-ray
is close to the central clump. In section~9 we investigate the 
substructure in more detail.

Towards the edges of the observed field the noise in the reconstruction 
increases rapidly, and therefore the significance of the structures at the 
edge of the field is low. 

The reconstruction after increasing the phase of the distortion by $\pi/2$ is
shown in figure~\ref{massrec}b. As this corresponds to the 
imaginary part of the mass surface density, no signal should be present if 
the observed distortion is due to gravitational lensing. We find no 
significant large scale structures in this map.

\section{Redshift distribution}

To relate the observed lensing signal to a measurement of the mass one has 
to estimate the mean critical surface density
$$\Sigma_{\rm crit}= c^2(4\pi G D_l \beta)^{-1},$$
where $D_l$ is the angular diameter distance to the lens and
$$\beta=\left\langle{\rm max}\left(0, D_{ls}/D_s\right)\right\rangle,$$
where $D_{s}$ and $D_{ls}$ are the angular diameter distances from observer to
the source and lens to the source respectively. The angular
diameter distance to MS~1054-03 is 1.96~$h_{50}^{-1}$~Gpc, which yields 
$\Sigma_{\rm crit}=845\beta^{-1} h_{50} \surfsun$ (using $\Omega_m=0.3$, and
$\Omega_\Lambda=0$).

The question that remains is the value of $\langle\beta\rangle$. The
uncertainty in the redshift distribution of the sources has hampered
the mass estimates from previous lensing studies of high redshift
clusters of galaxies (LK97; Clowe et al. 1998).

Down to $I\sim 23$ the redshift distribution of galaxies is fairly well known 
from redshift surveys, but the galaxies used in weak lensing analyses are 
generally too faint to be included in spectroscopic surveys. However,
photometric redshift distributions derived from broad band photometry of the
Northern HDF (Fern\'{a}ndez-Soto, Lanzetta, \& Yahil 1999) and Southern
HDF (Chen et al. 1998) may be used to obtain an estimate of 
$\langle\beta\rangle$ as a function of apparent magnitude and colour. 
The photometric redshift distributions can be compared to our data,
because we use the same filters, and the catalogs are complete down to 
the magnitude limit of our weak lensing analysis $(F814W<26.5)$.
Figure~\ref{sigcol} shows the comparison of the weak lensing signal with
the predicted trends based on the HDFs in independent magnitude and colour 
bins.

\vbox{
\begin{center}
\leavevmode
\hbox{%
\epsfxsize=7.6cm
\epsffile[80 370 430 670]{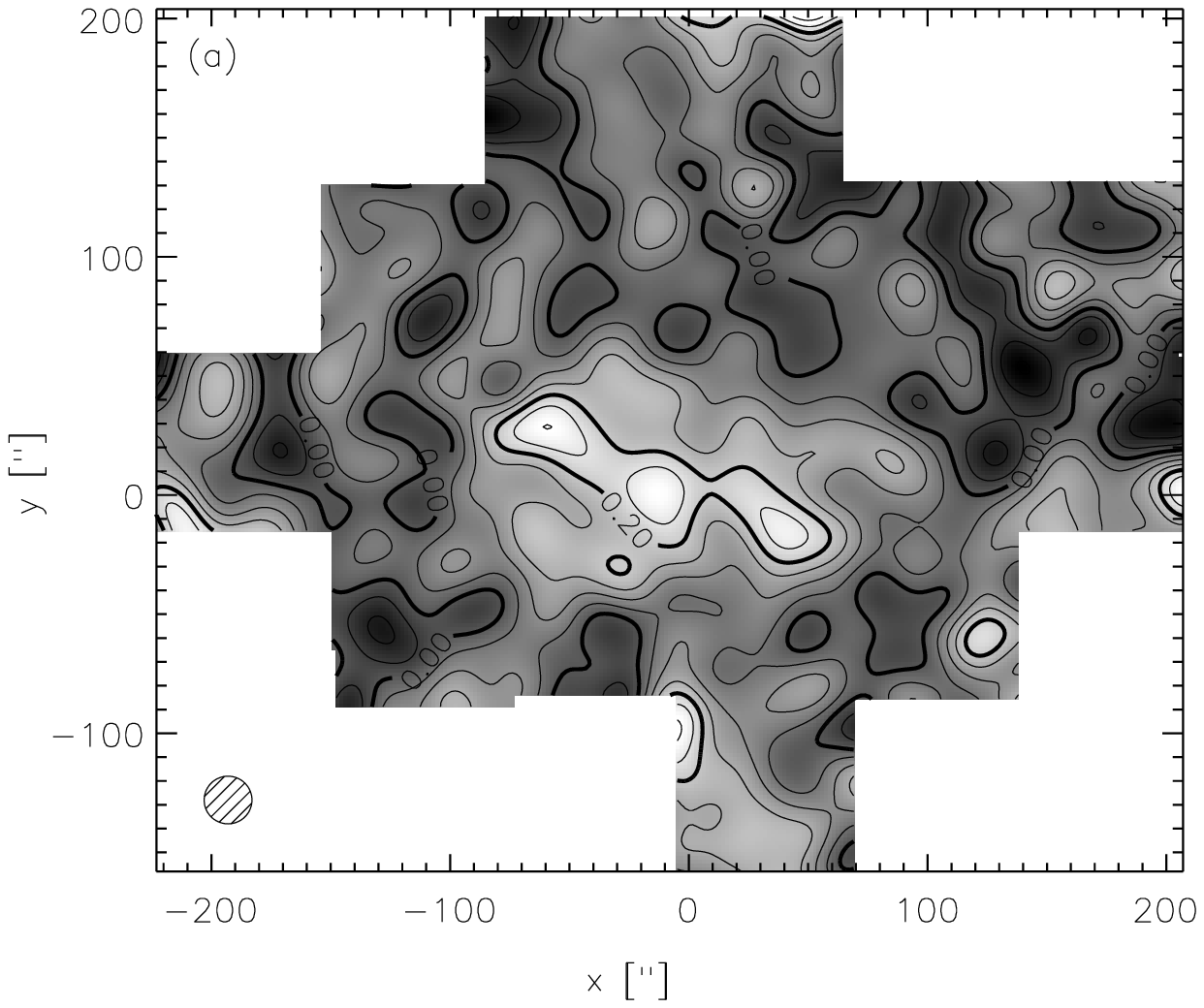}}
\hbox{%
\epsfxsize=7.6cm
\epsffile[80 370 430 670]{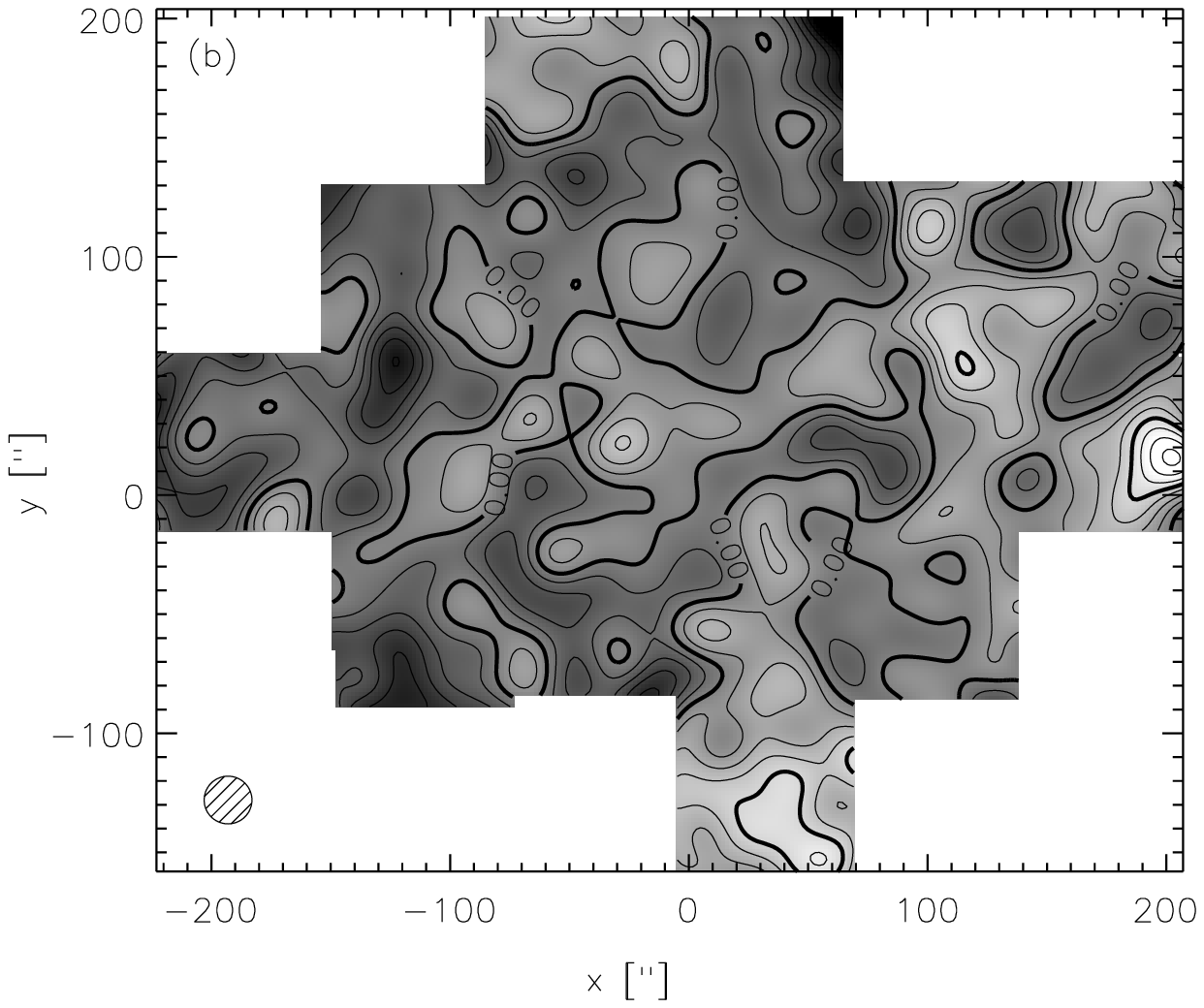}}
\begin{small}
\figcaption{(a) Reconstruction of the projected mass surface density. 
Towards the edges of the observations the signal-to-noise in the 
reconstruction decreases rapidly. The origin coincides with the assumed
cluster centre. The reconstruction shows that the mass distribution consists
of three distinct mass concentrations, similar to the light distribution.
(b) The reconstructed mass surface  density when the phase of the distortion 
is increased by $\pi/2$. As this projects out the imaginary part of the 
surface density, the signal should vanish if it is due to gravitational 
lensing. The mass distribution was smoothed using a Gaussian with a FWHM of 
20'' (indicated by the shaded circle). The interval between adjacent contours 
is 0.05 in $\kappa$.
\label{massrec}}
\end{small}
\end{center}}

Though there is some variation between the redshift distributions in the
Northern and Southern fields, the results from the weak lensing analysis 
agree well with the predictions, giving us confidence that the
calibration of the lensing signal using photometric redshifts works.

The average photometric redshift of $F814W<26.5$ galaxies in the HDF South 
is higher than it is in the Northern field. If this difference is 
representative of field to field variations the redshift distribution
the uncertainty in weak lensing mass estimates of high redshift clusters 
$(z\sim0.8)$ is significant: on the order of 10\% (a bootstrapping analysis 
of the photometric redshift catalog indicates that the relative uncertainty 
in $\langle\beta\rangle$ for each individual deep field is only 2\%).
For clusters at lower redshifts the introduced uncertainty is smaller.

The value of $\langle\beta\rangle$ does not only depend on the redshifts
of the sources, but it also depends on the cosmological parameters that
define the angular diameter distances. Figure~\ref{betacosm} shows
how the average value of $\beta$ depends on $\Omega_m$ and $\Omega_\Lambda$
for MS~1054. For a given value of the cosmological constant, the dependence 
on $\Omega_m$ is weak, but given $\Omega_m$, the average value for $\beta$ 
is more sensitive to the value of the cosmological constant. In an 
$\Omega_\Lambda$ dominated universe the value of $\beta$ is approximately 
20\% higher than in a low $\Omega_\Lambda$ universe, and as a result the 
estimated mass for MS~1054-03 would be 20\% lower. 

In our estimates of the lensing signal we weight the contribution of 
each source according to its error on the distortion (cf. appendix~A). 
Thus faint galaxies, which tend to be at higher redshifts are given less 
weight. Applying the same weight function (as a function of apparent
magnitude) to the HDF galaxies shows that the effective average $\beta$ is 
slightly (6\%) lower than the value indicated in figure~\ref{betacosm}. 
We use $\langle \beta \rangle=0.23$ $(\Omega_\Lambda=0)$ for our catalog of 
sources. Using $\Omega_m=0.3$ and $\Omega_\Lambda=0.7$ yields a higher value 
of 0.27 for $\langle \beta \rangle$.

\section{Mass estimates}

One of the advantages of gravitational lensing over other methods to estimate
masses of clusters is that no assumptions have to made about the dynamical
state or geometry. From the weak lensing signal we can directly estimate the
{\it projected} mass within some aperture, although this requires rather
good estimates for the redshift distribution of the background sources
(cf. section~6).

However, several effects complicate a precise measurement of the total
mass of a cluster. These are discussed in section~7.1. In section~7.2
we present our mass estimate for MS~1054-03. The result is compared
to other estimators in section~7.3.

\subsection{Complications in accurate mass determinations}

\subsubsection{Mass sheet degeneracy}

The mass sheet degeneracy (Gorenstein, Shapiro, \& Falco 1988)
reflects the fact that the observed distortion is invariant under
the transformation:

\begin{equation}
\kappa'=(1-\lambda)\kappa+\lambda.~\label{masdeg}
\end{equation}

Thus scaling the surface density, while adding a sheet of constant surface 
density, leaves the (observed) distortion unchanged.  Measuring the distortion
out to large radii, and arguing that the cluster surface density is neglegible
at large radii, solves the problem if no intervening sheet of matter is 
present. In the analysis presented here we assume that this is the case.

The degeneracy can be lifted by measuring the magnification of the sources 
relative to an unlensed population of sources (e.g. Broadhurst, Taylor, 
\& Peacock 1995). However, it is difficult to measure the magnification 
effect with sufficient accuracy to detect a mass sheet $\lambda$ of a few
percent.

\subsubsection{Measuring large distortions}

Simulations of sheared images show that the correction scheme we use to 
measure the shapes of the galaxies works well for relatively small distortions
$(\le 0.3)$. The KSB95 approach assumes that the distortion is small 
everywhere. The pre-seeing shear polarizability $P^\gamma$ is derived from 
the average over all objects, and hence it is not accurate in the regime 
where 

\vbox{
\begin{center}
\leavevmode
\hbox{%
\epsfxsize=7.2cm
\epsffile[18 144 575 650]{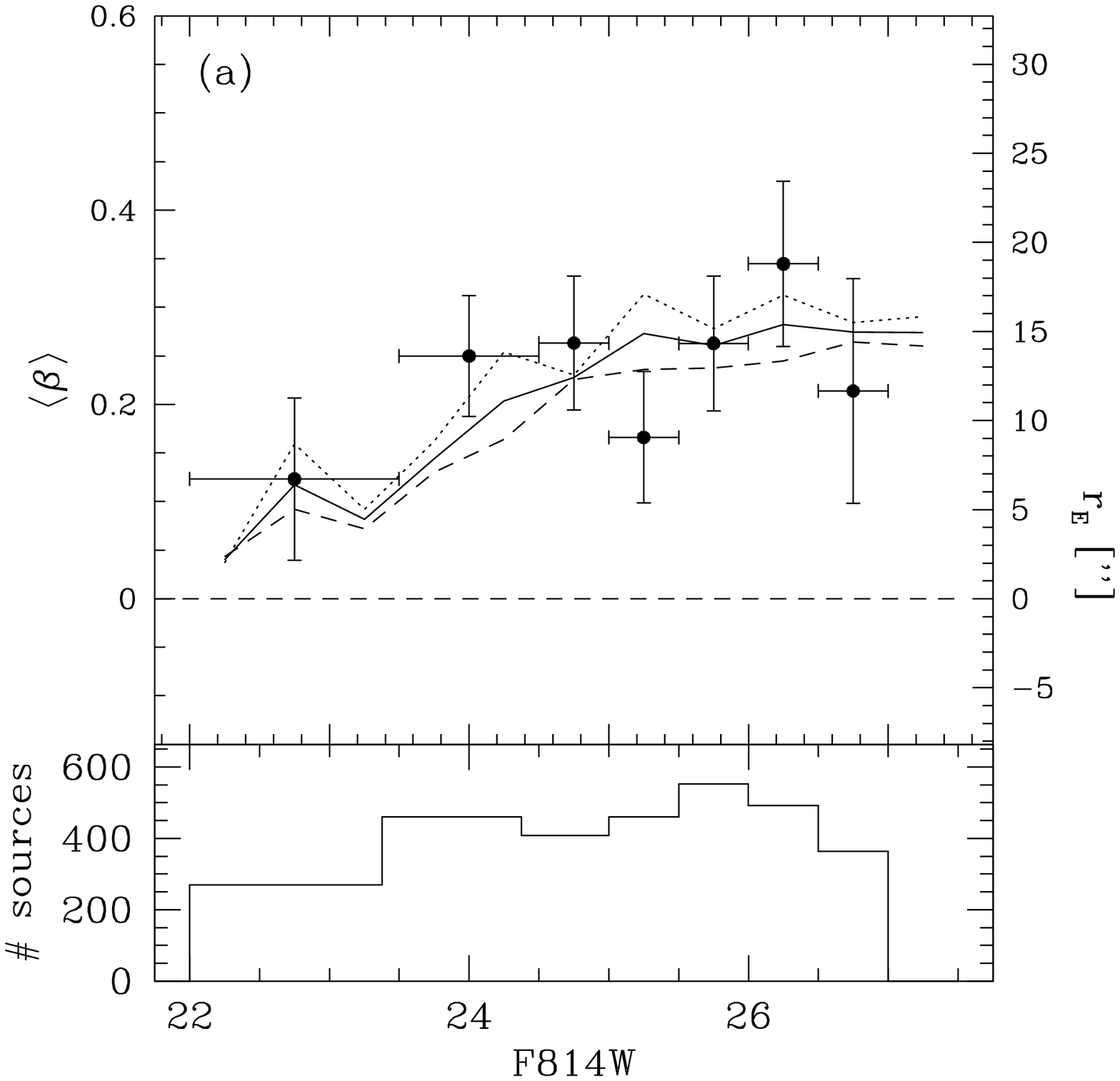}}
\hbox{%
\epsfxsize=7.2cm
\epsffile[18 144 575 680]{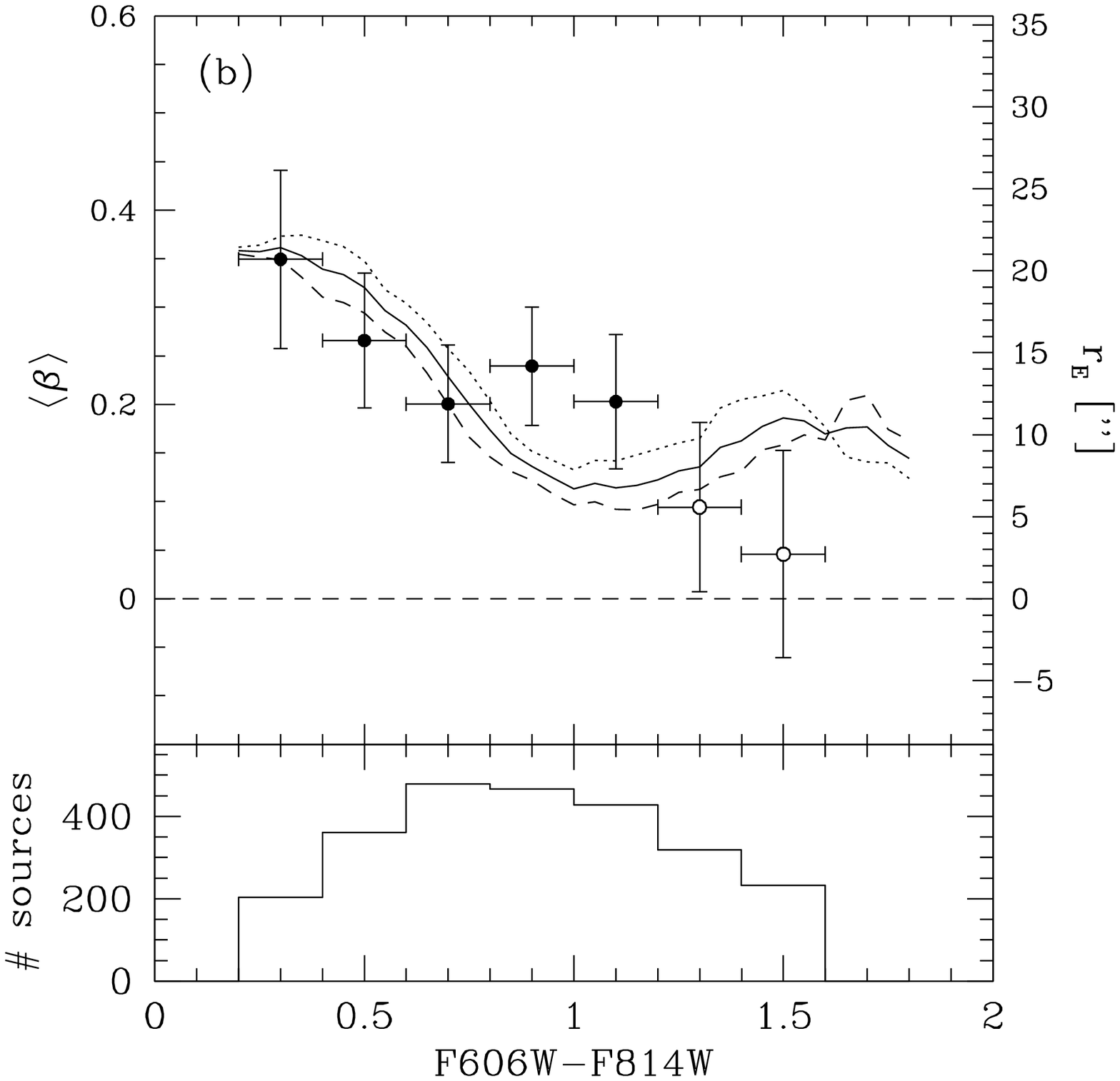}}
\begin{small}
\figcaption{(a) Comparison of the photometric redshift distribution
with the observed weak lensing signal. The solid line is the predicted
value of $\langle\beta\rangle$ as a function of apparent magnitude in 
the $F814W$ band, using the photometric redshift distribution of galaxies 
with $F606W-F814W < 1.6$ and $F814W<26.5$. The dashed line shows the results 
using only the photometric redshifts from the HDF-North, and the dotted
line gives the results for the Southern field. The variation in
the value of $\langle\beta\rangle$ obtained from the two deep fields
introduces an additional uncertainty in the weak lensing mass estimate. 
The points correspond to the  weak lensing signal, which is obtained by 
a SIS model fit to the data ($\kappa(r)=r_E/2r$, where $r_E$ corresponds to 
the Einstein radius.) to the data at radii larger than 75 arcseconds. The 
right $y-$axis indicates the vertical scale in $r_E$. (b) The comparison
as a function of colour of the background galaxies. We used galaxies with 
$22<F814W<26.5$. For the reddest bins (open circles), the contamination by 
cluster members is significant $(30-50\%)$, resulting in an underestimate 
of the lensing signal. The points in both figures are independent, and 
clearly show that the red galaxies carry a lower lensing signal and thus 
their average redshift is lower than that of the blue galaxies. The dependence
of the lensing signal on the apparent magnitude is quite weak for the faint 
galaxies. In both figures the lower panels show a histogram of the number of 
objects in each bin. The widths of the bins are indicated by the horizontal 
errorbars.
\label{sigcol}}
\end{small}
\end{center}}

\vbox{
\begin{center}
\leavevmode
\vspace{-0.5cm}
\hbox{%
\epsfxsize=8cm
\epsffile{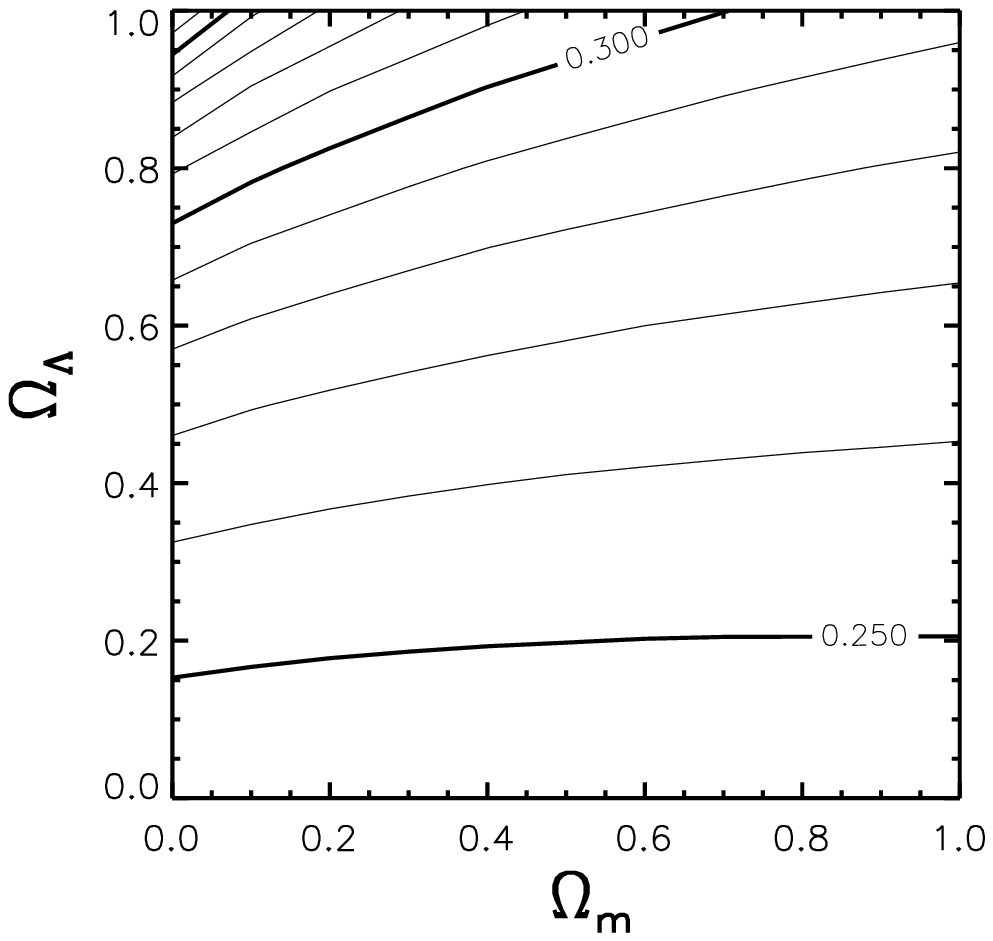}}
\begin{small}
\figcaption{Contour plot of $\langle\beta\rangle$ as a function of
$\Omega_m$ and $\Omega_\Lambda=1$, based on the photometric redshift 
distributions inferred from the HDF-N (Fern\'{a}ndez-Soto et al. 1999) 
and HDF-S (Chen et al. 1998). For a given value of $\Omega_\Lambda$,
the value of $\langle\beta\rangle$ depends only slightly on $\Omega_m$.
\label{betacosm}}
\end{small}
\end{center}}

\noindent the distortions are large. When the average distortion is 0.4, the 
method already underestimates the distortion by 10\%, and the discrepancy 
increases with increasing distortion. Other methods, like that proposed 
by Kuijken (1999), are more accurate when the distortion is large.

Due to the high redshift of MS~1054-03 the distortions are relatively
small (cf. figure~\ref{gtall}), and the KSB95 still works. However, 
placing the cluster at $z=0.2-0.4$ would produce much larger distortions, 
and the effect mentioned here has to be corrected for.

\subsubsection{Converting $g$ to $\gamma$}

The data provide a measure of the distortion $g$, which is related 
to the shear by $\gamma=(1-\kappa) g$. In the inner regions of the
cluster, where $\kappa$ is significant, the factor $(1-\kappa)$ is important.

\subsubsection{Broad redshift distribution}

Another effect is related to the redshift distribution of the 
sources. When computing the average distortion, we use an average value 
for $\beta$ for the sources, i.e. we assume that the redshift distribution 
can be approximated by a sheet of sources at redshift corresponding to the 
average $\beta$. As we show now, this results in an overestimate of the 
lensing signal.

For an individual galaxy the distortion is
\begin{equation}
g=\frac{\gamma}{1-\kappa}=\frac{\beta_s \gamma_\infty}
{1-\beta_s\kappa_\infty},\label{eqdistort}
\end{equation}
\noindent where $\kappa_\infty$ and $\gamma_\infty$ correspond to the 
dimensionless surface density and shear for a source at infinite redshift, 
and  $\beta_s=\beta/\beta_\infty$. If we adopt a common value of $\beta$ 
for all background galaxies, we estimate the distortion as 

\begin{equation}
\langle \tilde{g} \rangle =\frac{\langle\beta_s\rangle \gamma_\infty}
{1-\langle\beta_s\rangle \kappa_\infty}\label{g_wrong},
\end{equation}

\noindent whereas the true average distortion is in fact given by
the average of equation~\ref{eqdistort}:

\begin{equation}
\langle \hat{g} \rangle =\left\langle{\frac{\beta_s \gamma_\infty}
{1-\beta_s \kappa_\infty}}\right\rangle\label{g_ok}.
\end{equation}

\noindent The ratio $\hat g/\tilde g$ indicates by what
factor the shear $\gamma$ is overestimated. To first order
it can be approximated by:

\begin{equation}
\frac{\langle\hat{g}\rangle}{\langle\tilde{g}\rangle}
\approx 1+ \left(\frac{\langle\beta_s^2\rangle}{\langle\beta_s\rangle^2}-1\right)\kappa
= 1+ \left(\frac{\langle\beta^2\rangle}{\langle\beta\rangle^2}-1\right)\kappa.
\label{corkap}
\end{equation}

This result was also derived by Seitz \& Schneider (1997) in their
analysis of non-linear mass reconstructions. Using the photometric redshift
distributions for the HDFs, we find that for MS~1054-03, one 
would overestimate the shear by a factor $(1+0.62\kappa)$:
this represents a 7\% effect at a radius of 500 $h_{50}^{-1}$ kpc.
The size of the effect increases with lens redshift as is demonstrated in 
figure~\ref{corzdist}. 

\vbox{
\begin{center}
\leavevmode
\hbox{%
\epsfxsize=7cm
\epsffile{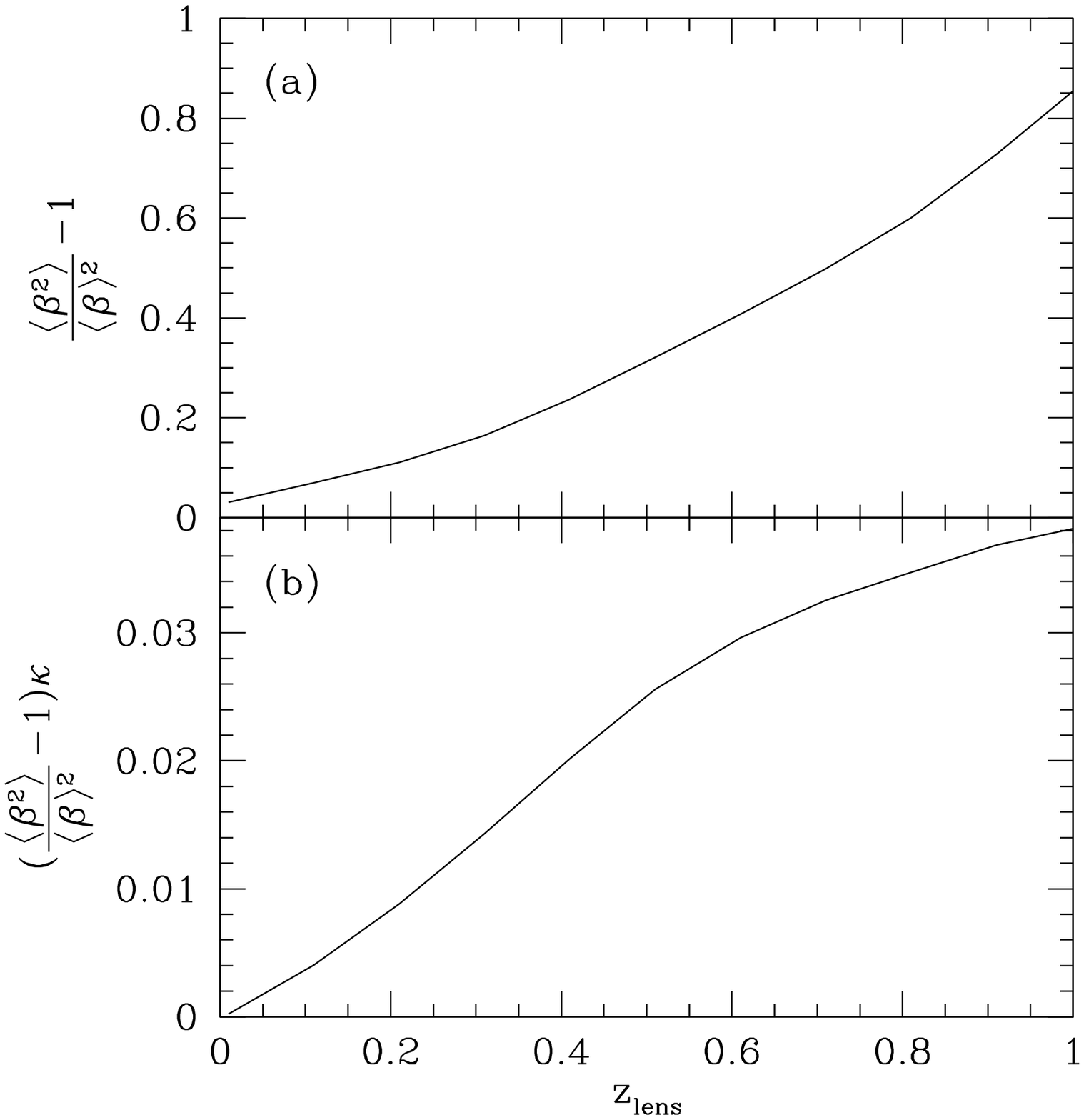}}
\begin{small}
\figcaption{(a) Plot of $(\langle\beta^2\rangle/\langle\beta\rangle^2-1)$ as
a function of lens redshift, based on the distribution of photometric redshifts
from the HDF-N and HDF-S. To produce this figure we used galaxies with
$F606W-F814W<1.6$ and  $F814W<26.5$. The effect increases with lens redshift. 
(b) Plot of $(\langle\beta^2\rangle/\langle\beta\rangle^2-1)\kappa$ as 
a function of lens redshift. We used the value of $\kappa$ at a distance
of 0.5$h_{50}^{-1}$ Mpc for a cluster with an isothermal mass distribution
and a velocity dispersion of 1000 km/s. For MS~1054-03 the shear at this 
distance from the cluster centre would be overestimated by 7\%.
\label{corzdist}}
\end{small}
\end{center}}

\subsubsection{Magnification}

A further source of error arises from the fact that lensed images of 
background sources are not only distorted, but also magnified. 
Their total flux is increased by a factor $\mu=((1-\kappa)^2-\gamma^2)^{-1}$. 
As a result the average redshift of the sources increases. The change in 
$\beta$ is approximately $2.2\kappa{\rm d}\beta/{\rm d}m$, which is small. 

\bigskip
All mass estimates presented below have been corrected for the effects
listed above. Since they are proportional to $\kappa$, they mainly affect
the inner regions of the cluster.

\subsection{Mass of MS~1054-03}

To obtain an estimate of the mass of MS~1054, we fit a simple mass model 
to the data. It is straightforward to include corrections for the effects 
mentioned in section~7.1.1 into the model. However, the result depends on the 
assumed radial surface density profile, and given the complex mass 
distribution of MS~1054, fitting a simple model might not yield the best 
mass estimate. Nevertheless, it should give a fair indication of the mass 
(within a factor of 1.5, depending on the radial profile). We fitted a 
singular isothermal sphere model $(\kappa=r_E/2r)$ to the tangential 
distortion at radii larger than 75''. We found a value of 
$r_E=11\farcs 5\pm 1\farcs 4$, which corresponds to a 
velocity dispersion $\sigma=1365\pm84$ km/s.

Another approach is often referred to as aperture mass densitometry. The 
shear can be related directly to a density contrast. We will use the 
statistic of Clowe et al. (1998), which is related to the $\zeta$ statistic 
of Fahlman et al. (1994):

\begin{equation}
\zeta_c(r_1)=2 \int \limits_{r_1}^{r_2} d\ln r \langle \gamma_T \rangle +
2(1-\frac{r_2^2}{r_{\rm max}^2})^{-1} \int \limits_{r_2}^{r_{\rm max}} 
d\ln r \langle \gamma_T \rangle\label{eqzetac}.
\end{equation}

$\zeta_c$ yields the mean dimensionless surface density interior to 
$r_1$ relative to the mean surface density in the annulus from $r_2$ to 
$r_{\rm max}$:

\begin{equation}
\zeta_c(r_1)=\bar\kappa(r'<r_1)-\bar\kappa(r_2<r'<r_{\rm max}).
\end{equation}

\noindent It shows that one can measure the average surface density
within $r_1$ up to a constant if the shear were observable.
We used the two-dimensional mass reconstruction, to obtain an estimate of 
the shear from the observed distortion.

\vbox{
\begin{center}
\leavevmode
\hbox{%
\epsfxsize=8cm
\epsffile{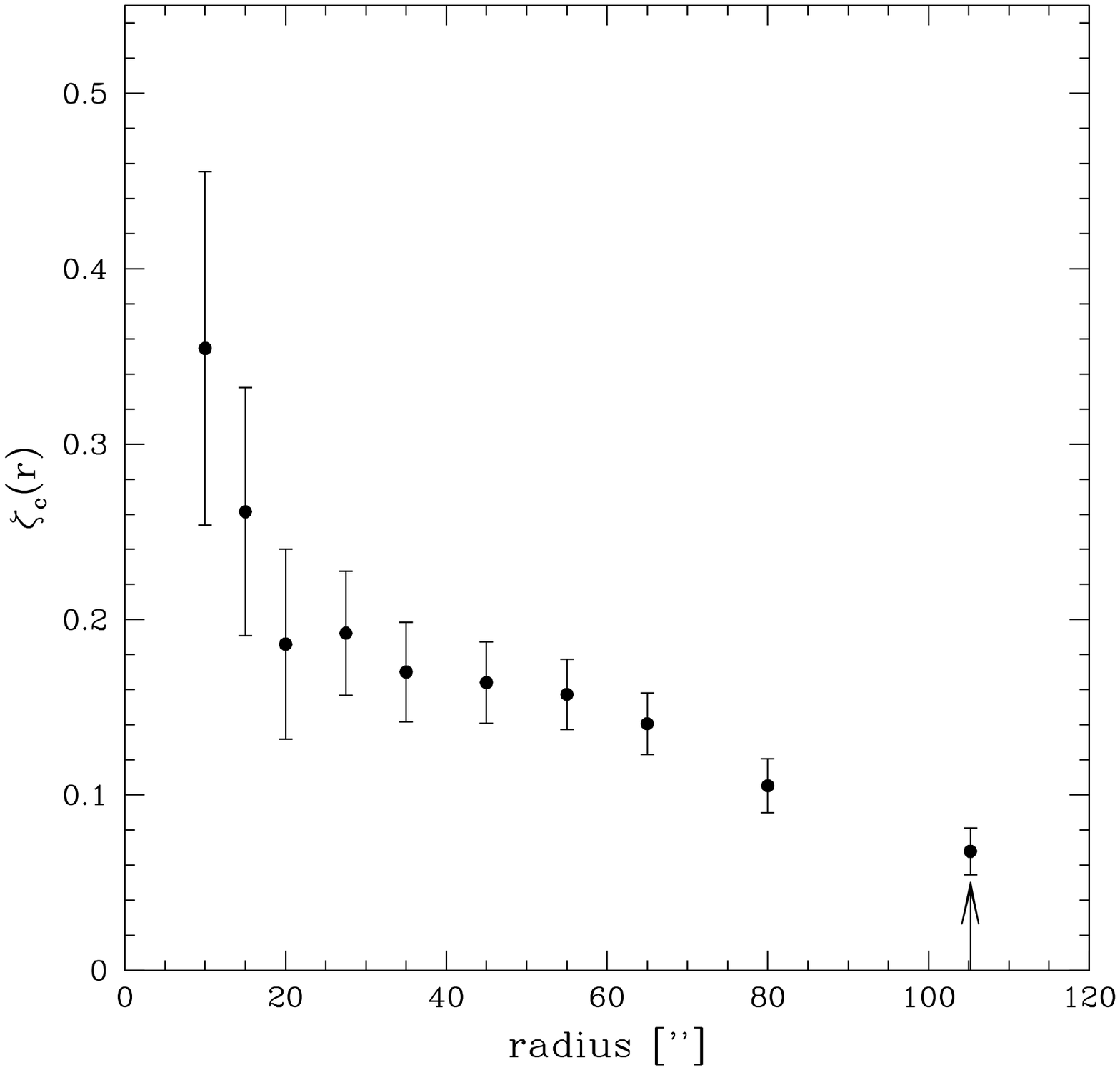}}
\begin{small}
\figcaption{Plot of the $\zeta_c$ statistic as a function of radius from the 
assumed cluster centre. The values for $\zeta_c$ were calculated using 
the sample of source galaxies (cf. table~\ref{table1}), and using 
$r_2=105.2''$ (1~$h_{50}^{-1}$ Mpc) and $r_{\rm max}=200''$. At small radii 
it is difficult to correct for the effects mentioned in 7.1.1, and as a result
these points are overestimates of the true values. The arrow indicates 
a radius of 1~$h_{50}^{-1}$ Mpc.
\label{zeta}}
\end{small}
\end{center}}

\begin{table*}
\begin{center}
\begin{small}
\begin{tabular}{ccccccc}
\hline
\hline
 (1) & (2) & (3) & (4) & (5) & (6) & (7) \\
$\Omega_m$  & $\Omega_\Lambda$ & 1~$h_{50}^{-1}$~Mpc & $\langle\beta\rangle$ & 
$\Sigma_{\rm crit}$ & $M(<1h_{50}^{-1} {\rm Mpc})$ & $\sigma$ \\
	    &           &       ['']	      & 		      &
$[\beta^{-1} h_{50} \surfsun]$ & $[10^{15}h_{50}^{-1}{\rm M}_\odot]$ & [km/s] \\
\hline
0.3       & 0.0 & 105.2	& 0.23 & 845 & $1.25 \pm 0.17$ & $1311^{+83}_{-89}$	\\
0.3       & 0.7	& 93.8	& 0.27 & 754 & $1.07 \pm 0.12$ & $1215^{+63}_{-67}$	\\
1.0       & 0.0	& 120.6	& 0.23 & 969 & $1.38 \pm 0.19$ & $1379^{+89}_{-97}$	\\
\hline
\hline
\end{tabular}
\caption{Mass estimates for various choices of $\Omega_m$ (1) 
and $\Omega_\Lambda$ (2); (3) angular scale of 1~$h_{50}^{-1}$~Mpc in 
arcseconds; (4) average $\beta$ for our catalog; (5) critical surface density;
(6) mass within an aperture of 1~$h_{50}^{-1}$~Mpc radius; (7) corresponding 
velocity dispersion if the mass distribution were isothermal.
\label{table2}}
\end{small}
\end{center}
\end{table*}

Figure~\ref{zeta} presents the measurement of $\zeta_c(r)$ using the
sample of source galaxies as defined in table~\ref{table1}
(cf. Fig.~\ref{gtall}), and taking $r_{\rm max}=200''$. At large
radii from the cluster centre we cannot average the data on complete
circles, and therefore we used equation~\ref{gtang} to estimate
the average tangential shear. The results show that the surface density in 
the centre is quite high, making it difficult to accurately correct the 
measurements at small radii for the effects listed in section~7.1.1. 

We find $\zeta_c=0.070\pm 0.014$ at a radius of 1~$h_{50}^{-1}$ Mpc
(indicated by the arrow). This yields a lower limit on the mass within this 
aperture of $(8.0\pm1.7)\times 10^{14}~h_{50}^{-1}~{\rm M}_\odot$. 
To obtain the total mass, we need to estimate the average surface density 
in the outer annulus. To this end we use the results of the SIS model fit.
This gives $0.038\pm0.005$ for the average dimensionless surface density in 
the outer annulus. Thus we find that the total mass within an aperture of 
1~$h_{50}^{-1}$ Mpc is $(1.2\pm0.2)\times 10^{15}~h_{50}^{-1}~{\rm M}_\odot$. 
If the mass distribution were isothermal, this mass estimate would correspond
to a velocity dispersion of $1311^{+83}_{-89}$ km/s. 

Table~\ref{table2} lists the mass estimates for different cosmologies. It 
shows that the mass estimate for MS~1054-03 decreases for an increasing 
value of the cosmological constant. 

The uncertainties in the mass estimates presented in this section only
reflect the contribution from the intrinsic ellipticities of the source 
galaxies. As shown in section~6, the uncertainty in the redshift
distribution of the sources may contribute an additional 10\% systematic
uncertainty.

\subsection{Comparison with other estimates}

For some time MS~1054-03 was the only very massive, X-ray selected,
high redshift cluster known, and therefore it has been studied
extensively. 

LK97 were the first to detect a weak lensing signal for this cluster. The 
tangential distortion presented in LK97 is about a factor 1.5 higher than the 
signal from our HST observations (using galaxies with $21.5<F814W<25.5$). 
The difference in the signals might be due to the large corrections for the 
circularization by the PSF in the ground based observations of LK97.
Furthermore, their mass estimate was hampered by the lack of knowledge of the 
redshift distribution of the faint galaxies.

Donahue et al. (1998) measured a high X-ray temperature of $12.3^{+3.1}_{-2.2}$
keV from ASCA observations, which corresponds to a velocity dispersion of 
$1400\pm170$ km/s (90\% confidence limits) under the assumption of an 
isothermal mass distribution. This estimate is in excellent agreement with 
our weak lensing mass estimate. However, the agreement is somewhat surprising:
the cluster is clearly not relaxed, and one might expect that the gas is shock
heated as the three massive clumps are merging. 

The dynamical velocity dispersion was measured by van Dokkum (1999), who
finds a value of $1150\pm90$ km/s based on 81 galaxies. Earlier reported
measurements based on smaller samples (Donahue et al. 1998; Tran et al. 1999) 
are consistent with this result.

All the evidence gathered so far indicates that MS~1054 is a very massive 
cluster, and using a Press-Schechter analysis one can estimate the likelihood 
of finding such a cluster in the EMSS catalog. Based on such analyses, LK97, 
and Donahue et al. (1998) have argued that the presence 
of MS~1054 in the EMSS catalog is hard to reconcile with $\Omega_m=1$. 
However, in an open or $\Omega_\Lambda$ dominated universe, the likelihood of 
finding massive high redshift clusters is much greater 
(e.g. Bahcall \& Fan 1998). If the primordial density fluctuations are 
non-gaussian, the presence of MS~1054-03 in the EMSS catalog does not 
necessarily exclude a high density universe (Willick 1999). 

\vbox{
\begin{center}
\leavevmode
\hbox{%
\epsfxsize=8cm
\epsffile{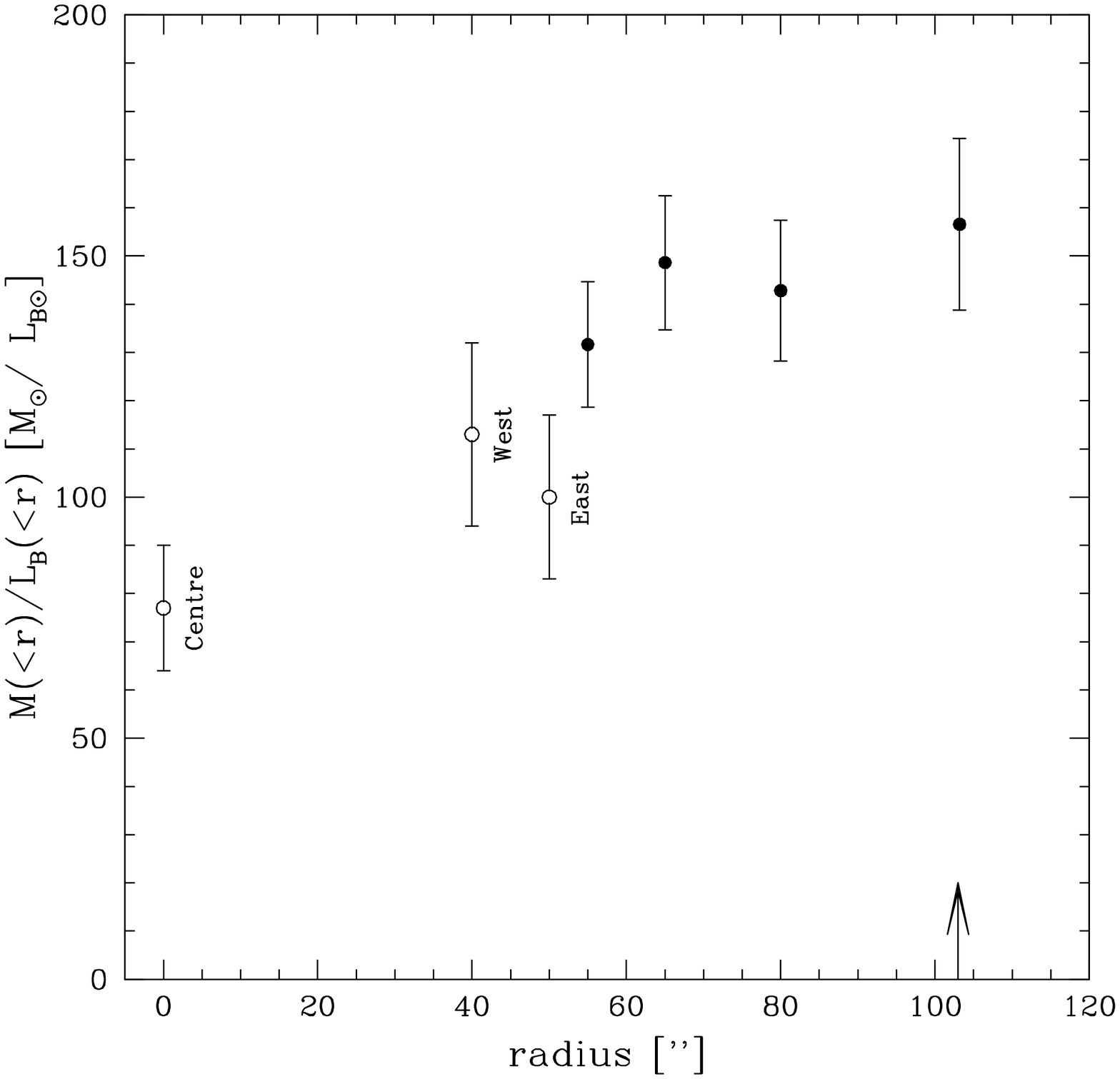}}
\begin{small}
\figcaption{Average mass-to-light ratio within an aperture for 
radii larger than 50 arcseconds from the cluster centre (filled circles). 
The mass in the aperture was estimated using equation~\ref{eqzetac} and
assuming that the average dimensionless surface density in the
outer annulus $\kappa(105{\farcs}2<r<200'')=0.038\pm0.005$.
The points are correlated and the errorbars only reflect the
uncertainty in the measurement of the distortion. The open
circles denote the observed mass-to-light ratios of the three
clumps in the mass distribution, using the results listed
in table~\ref{tableclump}. The arrow indicates a radius of 
1~$h_{50}^{-1}$ Mpc.
\label{moverlbar}}
\end{small}
\end{center}}

\begin{table*}
\begin{center}
\begin{tabular}{ccccc}
\hline
\hline
 (1) & (2) & (3) & (4) & (5) \\
$\Omega_m$  & $\Omega_\Lambda$ & $L_B$ & $M/L_B$ & $M/L_B~(z=0)$ \\
  &  & $[10^{13}~h_{50}^{-2} L_{B\odot}]$ & $[h_{50} {\rm M}_\odot/{\rm L}_{B\odot}]$ & 
$[h_{50} {\rm M}_\odot/{\rm L}_{B\odot}]$ \\
\hline
0.3       & 0.0 & 1.0  & $124\pm17$ & $269\pm37$\\
0.3       & 0.7	& 1.2  & $90\pm10$   & $225\pm25$\\
1.0       & 0.0	& 0.85 & $163\pm22$ & $299\pm41$\\
\hline
\hline
\end{tabular}
\begin{small}
\caption{Average mass-to-light ratio within an aperture with a 1~$h_{50}^{-1}$ 
Mpc radius for different choices of the cosmological parameters (columns (1) 
and (2)); (3) Total light within the aperture; (4) observed average 
mass-to-light ratio in the passband corrected $B$ band; (5) mass-to-light 
ratio in the $B$ band, corrected for luminosity evolution as inferred from 
the evolution of the fundamental plane for the various choices of $\Omega_m$ 
and $\Omega_\Lambda$ (van Dokkum et al. 1998).
\label{tableml}}
\end{small}
\end{center}
\end{table*}

\begin{table*}[b!]
\begin{center}
\begin{small}
\begin{tabular}{lcccccc}
\hline
\hline
(1) & (2) & (3) & (4) & (5) & (6) & (7) \\
Clump  	&$\sigma_{SIS}$ & $\zeta_c(26'')$ & $\bar\kappa(100''<r<150'')$ &
	M ($<250~h_{50}^{-1}$ kpc) & $L_B$ & $M/L_B$ \\
       	& [km/s] & & & $[10^{14}h_{50}^{-1}{\rm M}_\odot]$ & $[10^{13}h_{50}^{-2}L_{B\odot}]$ &
	$[h_{50} {\rm M}_\odot/{\rm L}_{B\odot}]$ \\
\hline
East    & $702\pm75$ &	$0.184 \pm 0.040$ & $0.050 \pm 0.006$ & $1.7 \pm 0.3$ & 1.7 & $100 \pm 17$ \\   
Centre  & $623\pm76$ &	$0.194 \pm 0.037$ & $0.047 \pm 0.006$ & $1.7 \pm 0.3$ & 2.2 & $77  \pm 13$ \\	
West    & $628\pm80$ &	$0.181 \pm 0.037$ & $0.050 \pm 0.006$ & $1.7 \pm 0.3$ & 1.5 & $113 \pm 19$ \\
\hline
\hline
\end{tabular}
\caption{Mass estimates for the subclumps. (1) position of the clump; 
(2) velocity dispersion obtained by simultaneously fitting 3 singular 
isothermal spheres to the observed distortion; (3) the observed value of 
$\zeta_c$ at 250~$h_{50}^{-1}$ kpc; (4) the average dimensionless surface 
density in the control annulus ($\bar\kappa(100''<r<150''))$ assuming an 
isothermal mass distribution with $\sigma=1311\pm85$ km/s ; (5) total 
projected mass within an aperture of 250~$h_{50}^{-1}$ kpc radius; (6) 
total luminosity in an aperture of 250~$h_{50}^{-1}$ kpc radius; (7) 
average mass-to-light ratio in an aperture of 250~$h_{50}^{-1}$ kpc.
\label{tableclump}}
\end{small}
\end{center}
\end{table*}

\section{Mass-to-light ratio}

Combining measurement of the mass with the estimate of the total light
in the passband corrected $B$ band within an aperture of 1~$h_{50}^{-1}$ 
Mpc radius (cf. section~4) yields $(124\pm17) h_{50} {\rm M}_\odot/
{\rm L}_{B\odot}$. This result has to be corrected for luminosity evolution
in order to compare it to values found for lower redshift clusters.
Studies of the fundamental plane of clusters of galaxies at various redshifts
have quantified the luminosity evolution (e.g. van Dokkum \& Franx 1996; 
Kelson et al. 1997; van Dokkum et al. 1998). The fundamental plane of 
MS~1054 was studied by van Dokkum et al. (1998). They found that the 
mass-to-light ratio of early type galaxies in  MS~1054-03 is $\sim 50\%$ lower
than the present day value. Under the assumption that the total luminosity of 
the cluster has changed by the same amount, the corrected mass-to-light ratio 
of MS~1054 is $269\pm37 h_{50} {\rm M}_\odot / {\rm L}_{B\odot}$.

The resulting mass-to-light ratio depends on the cosmological parameters. 
In table~\ref{tableml} the observed mass-to-light ratios in the passband 
corrected $B$ band are listed (column~(4)). Column~(5) lists the mass-to-light 
ratio corrected for luminosity evolution as determined from the evolution of 
the fundamental plane for MS~1054-03 for the various cosmologies considered
here (van Dokkum et al. 1998).

HFKS98 measured a mass-to-light ratio of $(175\pm26) h_{50} 
{\rm M}_\odot/{\rm L}_{B\odot}$ for Cl~1358+62, which changes to 
$(231\pm34) h_{50} {\rm M}_\odot/{\rm L}_{B\odot}$ after correction
for luminosity evolution. Carlberg, Yee, \& Ellingson (1997) studied a
sample of 16 rich clusters of galaxies, and they find a value of
$(134\pm9) h_{50} {\rm M}_\odot/{\rm L}_{r\odot}$ for the average mass-to-light
ratio. Using $B-r\sim 1.1$ as found for nearby early type galaxies 
(J{\o}rgensen, Franx, \& Kjaergaard 1995) this transforms to 
$(260\pm17) h_{50} {\rm M}_\odot/{\rm L}_{B\odot}$ in the $B$ band. 

This comparison indicates that the mass-to-light ratio of MS~1054 is 
similar to what is found from other studies, suggesting that the
range in cluster mass-to-light ratios is rather small. LK97 argued that 
MS~1054-03 might be a relatively dark cluster, but our analysis does not
confirm their results.

We also studied the radial dependence of the mass-to-light ratio using
equation~\ref{eqzetac} and an estimate for the average surface density in 
the control annulus. This allows us to estimate the average mass-to-light 
ratio $M(<r)/L_B(<r)$ as a function of radius. To avoid confusion with the
clumps in the mass distribution, we present the results for radii larger
than 50 arcseconds in figure~\ref{moverlbar} (solid points). The open circles 
indicate the mass-to-light ratios measured for the three clumps in the mass 
distribution, using the results listed in table~\ref{tableclump}. The 
results indicate that the light is more concentrated towards the centre
of the cluster than the mass.

\section{Substructure}

The data enable us to study the observed substructure in the cluster
mass distribution in detail. Figure~\ref{masscen}a shows the mass 
reconstruction of the central 3.3 by 2.7 arcminute of the HST mosaic. 
The origin coincides with the assumed cluster centre. Figure~\ref{masscen}b 
is the overlay of the reconstruction and the $F814W$ image. To allow a direct 
comparison of the mass reconstruction with the light distribution, we also 
present the overlay of the smoothed light distribution in 
figure~\ref{masscen}c. This shows that the projected mass distribution 
compares well with the light distribution.

Next we estimate the masses of the three clumps from the observed
distortions (see Table~\ref{tableclump}). We use two measures:

\begin{itemize}
\item[$(i)$]{a simultaneous fit of three isothermal spheres to the full 
distortion field;}
\item[$(ii)$]{the $\zeta_c$-statistic (equation~\ref{eqzetac}).}
\end{itemize}

A simultaneous fit of three SIS models to the observed distortion 
shows that the masses of the three clumps are similar, with corresponding
velocity dispersions of $600-700$ km/s. Comparison with the estimate
of the total cluster mass indicates that most of the mass is in
the three clumps. For the $\zeta_c$ statistic we need 
to estimate the average dimensionless surface density in a control annulus 
around each clump (we use $r_2=100''$ and $r_{\rm max}=150''$ as at these 
radii the effect of the substructure is expected to be small). To do so, 
we assume that the surface density profile at large radii is isothermal 
with a velocity dispersion of $\sigma=1311\pm85$ km/s as found above. The 
results are listed in Table~\ref{tableclump}.

The results from the $\zeta_c$-statistic correspond to the total projected 
mass in the aperture, which also includes mass from the other clumps. 
Therefore we consider the results of the fitting procedure (column~(1)) as 
the best estimates for the mass.

However, to estimate the average mass-to-light ratio of the clumps within a 
radius of 250~$h_{50}^{-1}$ kpc we do use the $\zeta_c$ statistic, as both
mass and light are measured in the same aperture. The mass-to-light ratios are
around $90 h_{50} {\rm M}_\odot/{\rm L}_{B\odot}$, somewhat lower 
than the average mass-to-light ratio within an 1~$h_{50}^{-1}$~Mpc radius 
aperture around the cluster centre.

New X-ray telescopes like Chandra or XMM will have the sensitivity 
to study the X-ray properties of high redshift clusters in detail. 
MS~1054-03 will be an interesting target as we know how the mass is 
distributed. Thus high resolution measurements of the X-ray emission and 
temperature gradients can provide important information about the 
thermodynamics of cluster formation.

\subsection{Relaxation timescale}

The redshift catalog of van Dokkum (1999) enables us to study the
velocity structure of the cluster. The galaxies more that 100'' west
of the centre show a velocity difference of about 800 km/s with
respect to the average velocity. However, their contribution to the
total velocity dispersion is small. The three clumps are located
closer to the centre, and cannot be separated in velocity. Therefore
the radial motion of the clumps appears to be small compared to the
velocity dispersion of the cluster.

Both outer clumps in the mass distribution are in projection 
500$h_{50}^{-1}$ kpc away from the central clump. If we assume 
that all motion is perpendicular to the line of sight, and using a 
relatively slow approach (500 km/s) we estimate the timescale for 
relaxation to be on the order of 1 Gyr. Therefore the cluster should 
be relaxed at a redshift of $z\sim 0.65$. If the physical separations of 
the clumps are larger, the relaxation time increases correspondingly.

\section{Galaxy-galaxy lensing}

To study the mass distribution of MS~1054-03, we concentrated on the 
lensing signal on relatively large scales. However, the cluster galaxies
themselves introduce a small signal as well. Measuring this galaxy-galaxy 
lensing signal can be used to study the halo properties of field galaxies 
(e.g. Brainerd, Blandford, \& Smail 1996; Schneider \& Rix 1997, Hudson et al.
1998).

\vbox{
\begin{center}
\leavevmode
\hbox{%
\epsfxsize=7.5cm
\epsffile[85 330 440 690]{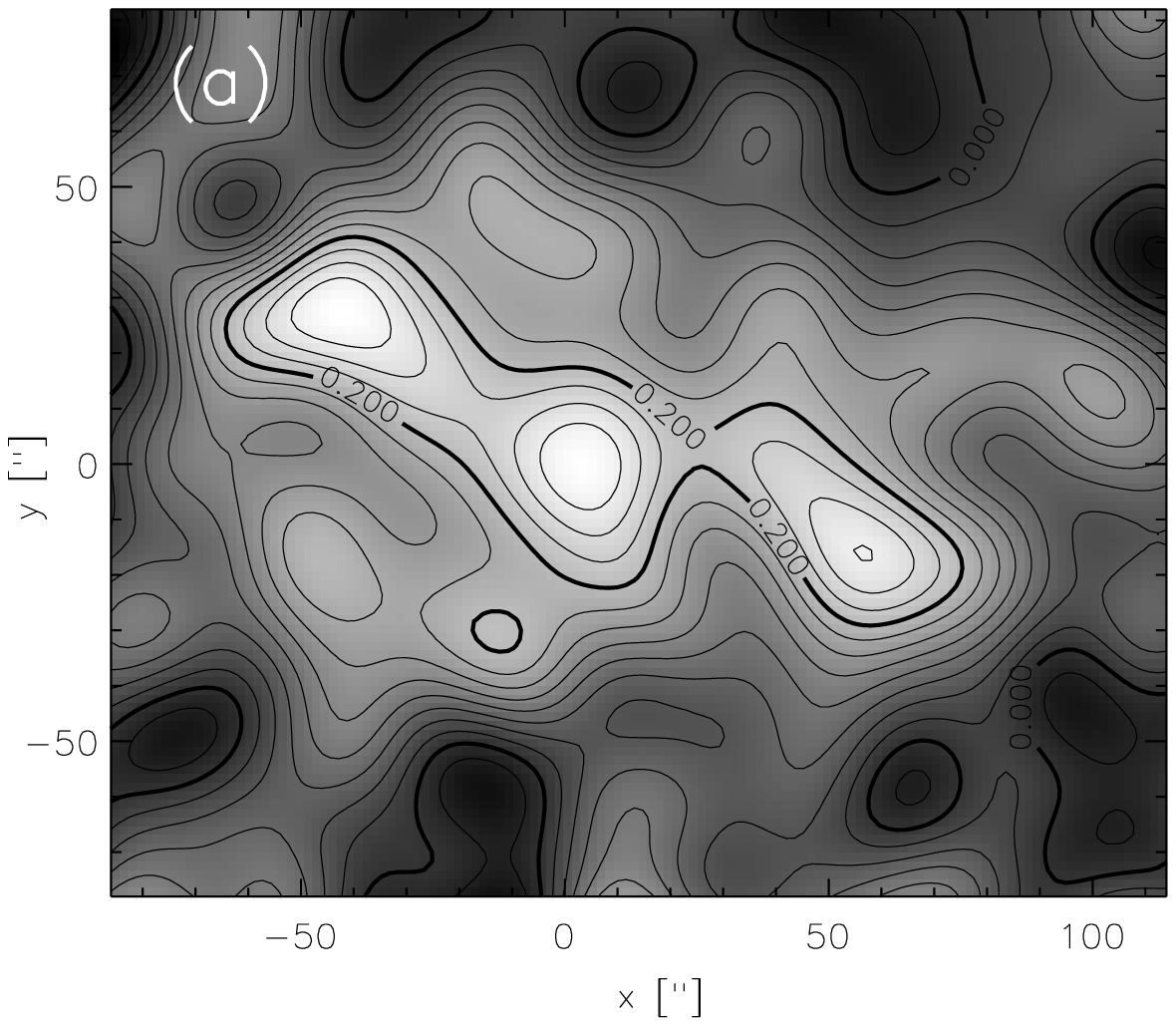}}
\hbox{%
\epsfxsize=7.5cm
\epsffile{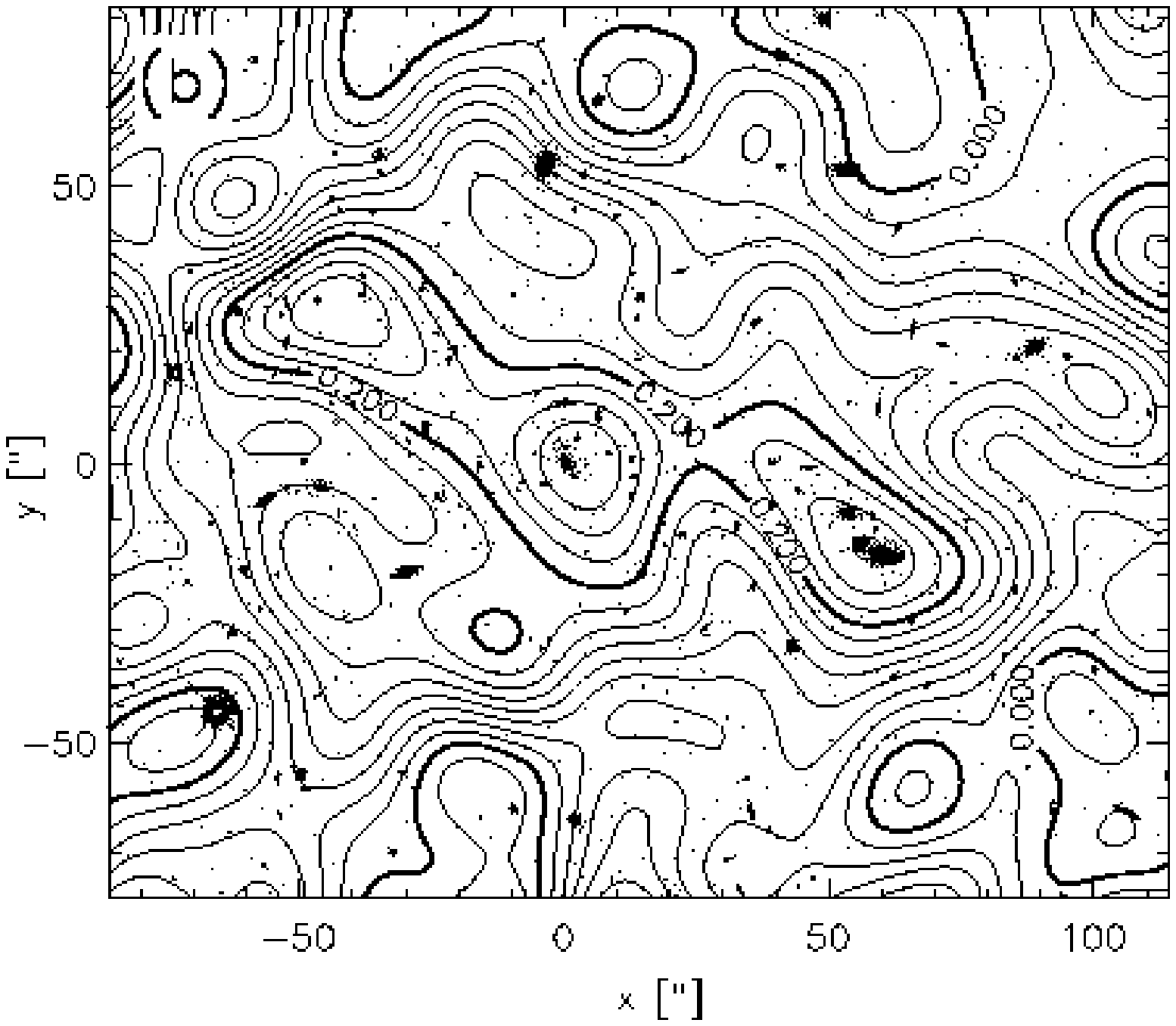}}
\hbox{%
\epsfxsize=7.5cm
\epsffile{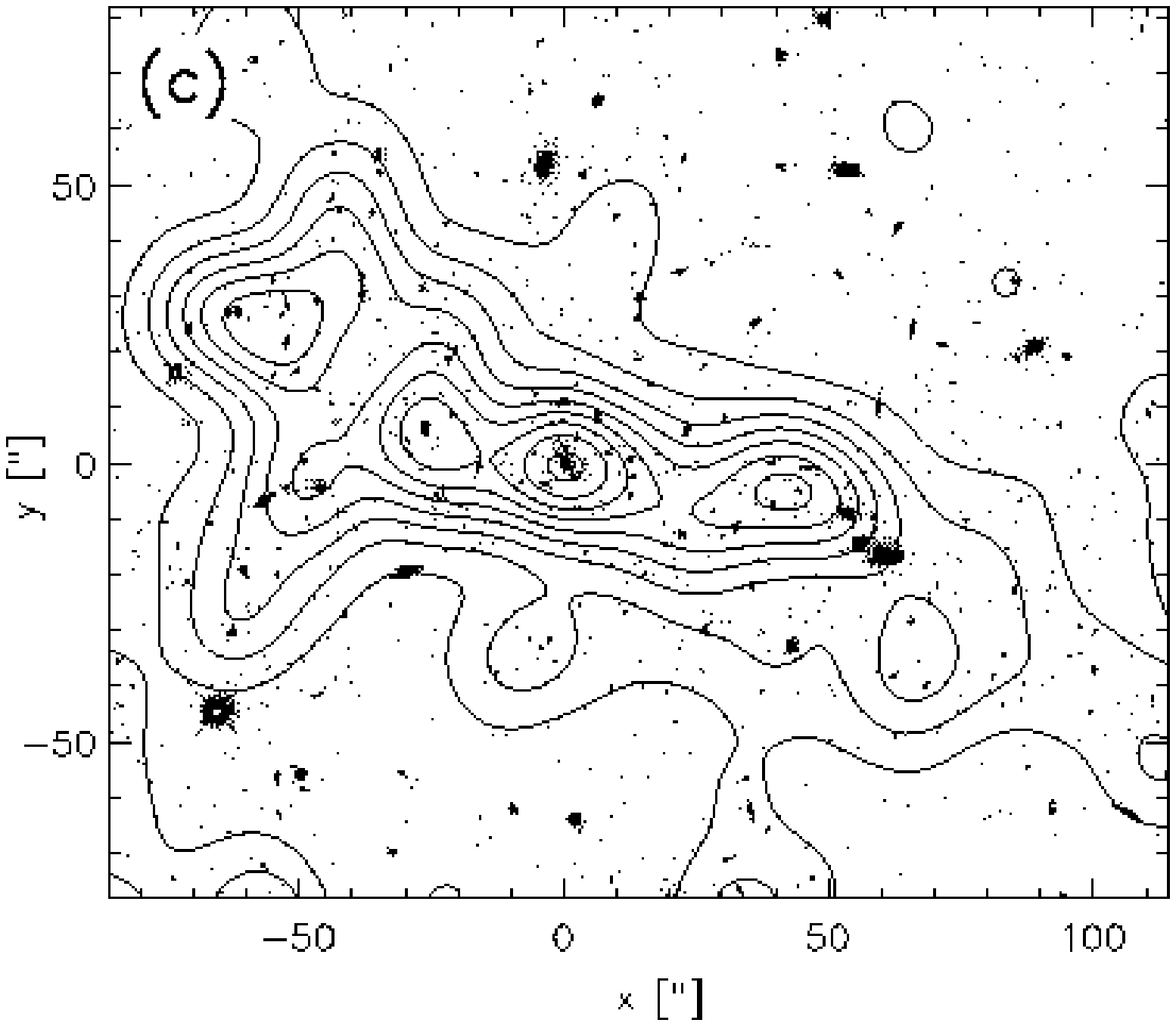}}
\begin{small}
\figcaption{(a) Gray scale image of the mass reconstruction of the
central 3.3 by 2.7 arcminute region of the HST mosaic; (b)
Overlay of the mass map on the $F814W$ image of MS~1054-03; (c)
Overlay of the smoothed light map and the $F814W$ image. All images
have been smoothed using a Gaussian filter with a FWHM of 20 arcsec.
The interval between adjacent contours is 0.025 in $\kappa$.
Comparison of the mass reconstruction with the light distribution 
shows that there is a fair agreement between the distributions of mass 
and light.
\label{masscen}}
\end{small}
\end{center}}

A robust method to examine the cluster galaxies is to measure the
average tangential distortion around an ensemble of cluster members.
The results are not very sensitive to the precise cluster mass distribution,
provided one confines the analysis to small radii. Due to the clustering of 
the lenses, the signal at large radii is dominated by the smooth cluster 
signal. When measuring the signal out to large radii, a careful removal of 
the cluster signal is required. We subtracted the contribution of the smooth 
cluster mass distribution from the observed distortion using the 
two-dimensional mass reconstruction, and computed the shear $\gamma$
taking the $(1-\kappa)$ correction into account. 

We used a sample of galaxies brighter than $F814=24.5$ $(\sim 0.05L_*)$, 
with colours close to the cluster colour-magnitude relation to study the 
masses of the cluster galaxies. To minimize the effect of the large scale 
cluster mass distribution, we excluded galaxies near the centres of the 
three clumps, resulting in a sample of 241 galaxies. In total, 4300 cluster 
galaxy-background galaxy pairs were used to measure the distortion.

Massive galaxies produce a larger distortion than less massive galaxies and
therefore we normalized the amplitudes of the signals from the various 
galaxies correspondingly. To do so, we assumed that the shear 
$\gamma\propto\sigma^2\propto\sqrt{L_B}$, i.e. we assumed a Faber-Jackson 
scaling relation. The results are presented in figure~\ref{galgal}.

A fit of a singular isothermal model to the data, yielded a detection
of the lensing signal at the 99.8\% confidence limit. We found for the 
Einstein radius of an $L_*$ galaxy (${\rm L}_{*B}=8\times 10^{10} 
h_{50}^{-2} {\rm L}_{B\odot}$ for MS~1054-03) a value of 
$r_{\rm E}=0.26\pm0.09$ arcseconds, which corresponds to a velocity dispersion 
of $\sigma=203\pm33$ km/s. The mass-to-light ratio within a radius of 
$20 h_{50}^{-1}$ kpc is ${\rm M}/{\rm L}_B=7.5\pm2.4 h_{50}
{\rm M}/{\rm L}_{B\odot}$. Correcting this result for luminosity evolution
yields ${\rm M}/{\rm L}_B=17\pm5 h_{50}{\rm M}/{\rm L}_{B\odot}$.

If we neglect the effect of the smooth cluster contribution, we find a 
slightly higher mass for the galaxies: $\sigma=212\pm32$ km/s. This  
indicates that the mass estimate determined using the direct averaging 
method is indeed robust. 

Natarayan \& Kneib (1998) studied the halo properties of the cluster galaxies
in AC~114. They used a maximum-likelihood analysis to constrain the mass and 
the extent of the galaxy halos. Constraining the extent of the halos of the 
cluster galaxies in MS~1054 is a delicate task which has not been attempted 
here. It requires a careful analysis of the smooth cluster mass distribution 
in a maximum likelihood analysis.

\section{Conclusions}

We have measured the weak gravitational lensing of faint, distant galaxies
by MS~1054-03, a rich cluster of galaxies at $z=0.83$. The data consist of 
a two-colour mosaic of 6 interlaced images taken with the WFPC2 camera on 
the Hubble Space Telescope.

The expected lensing signal is low for high redshift clusters, and 
most of the signal comes from small, faint galaxies. Compared to ground
based observations, these galaxies are much better resolved in HST 
observations. High number densities of sources are reached, and the correction
for the circularization by the PSF is much smaller. This enables us to
measure an accurate and well calibrated estimate of the lensing induced 
distortion of the faint background galaxies. 

Several effects complicate an accurate analysis when the distortions
or the surface density are non-neglegible (e.g. in the cluster cores). 
We do not measure the shear, but the distortion $g$. Neglecting the effect
of a non-zero surface density results in an overestimate of the lensing
signal. Furthermore, the method used (KSB95; LK97) underestimates the 
distortion when it is larger than $\sim 0.3$. Although we correct for
these effects, it is best to restrict the analysis to the weak lensing regime, 
where the convergence $\kappa$ is low, and the distortions are small.

\vbox{
\begin{center}
\leavevmode
\hbox{%
\epsfxsize=7.5cm
\epsffile{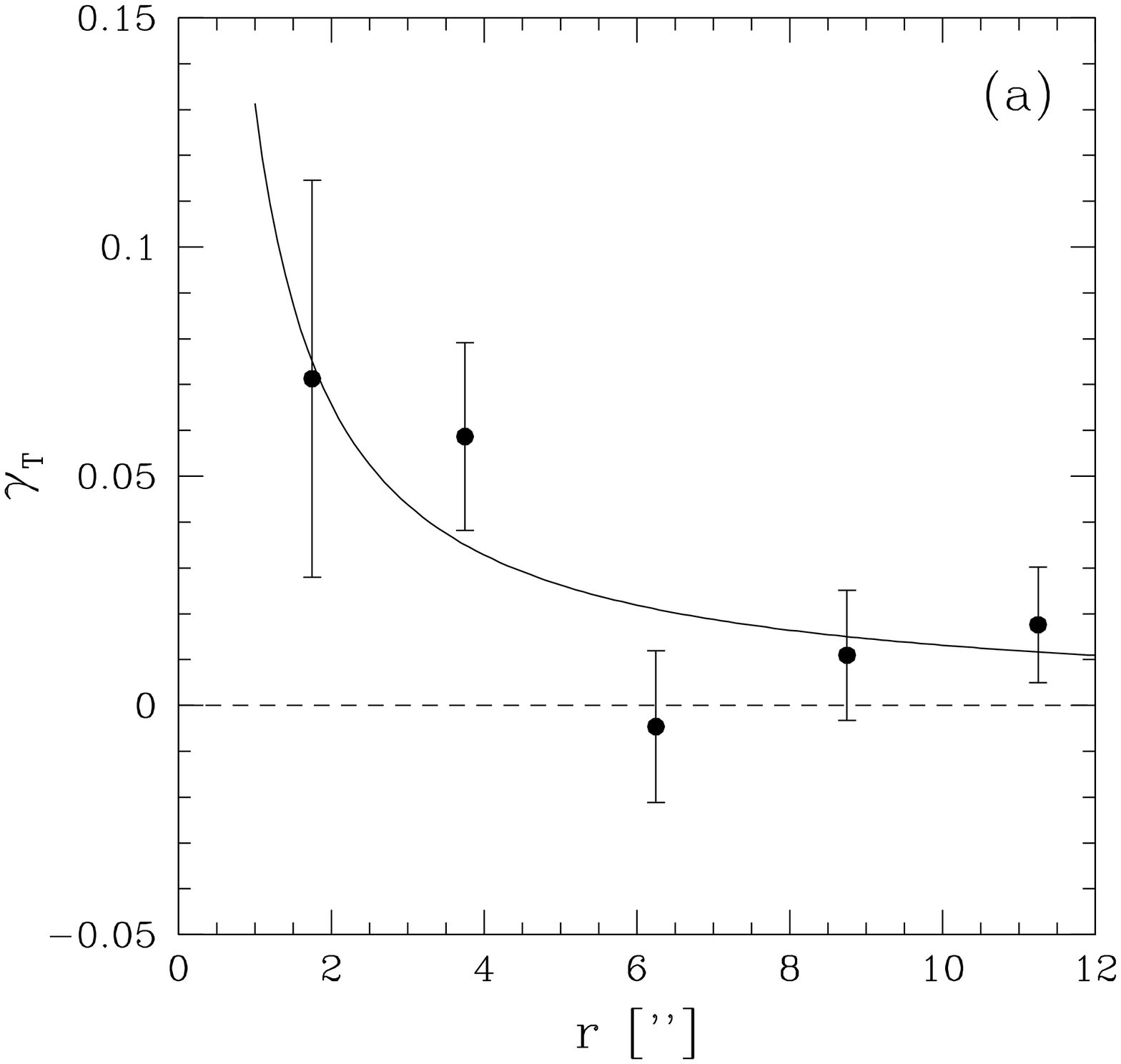}}
\hbox{%
\epsfxsize=7.5cm
\epsffile{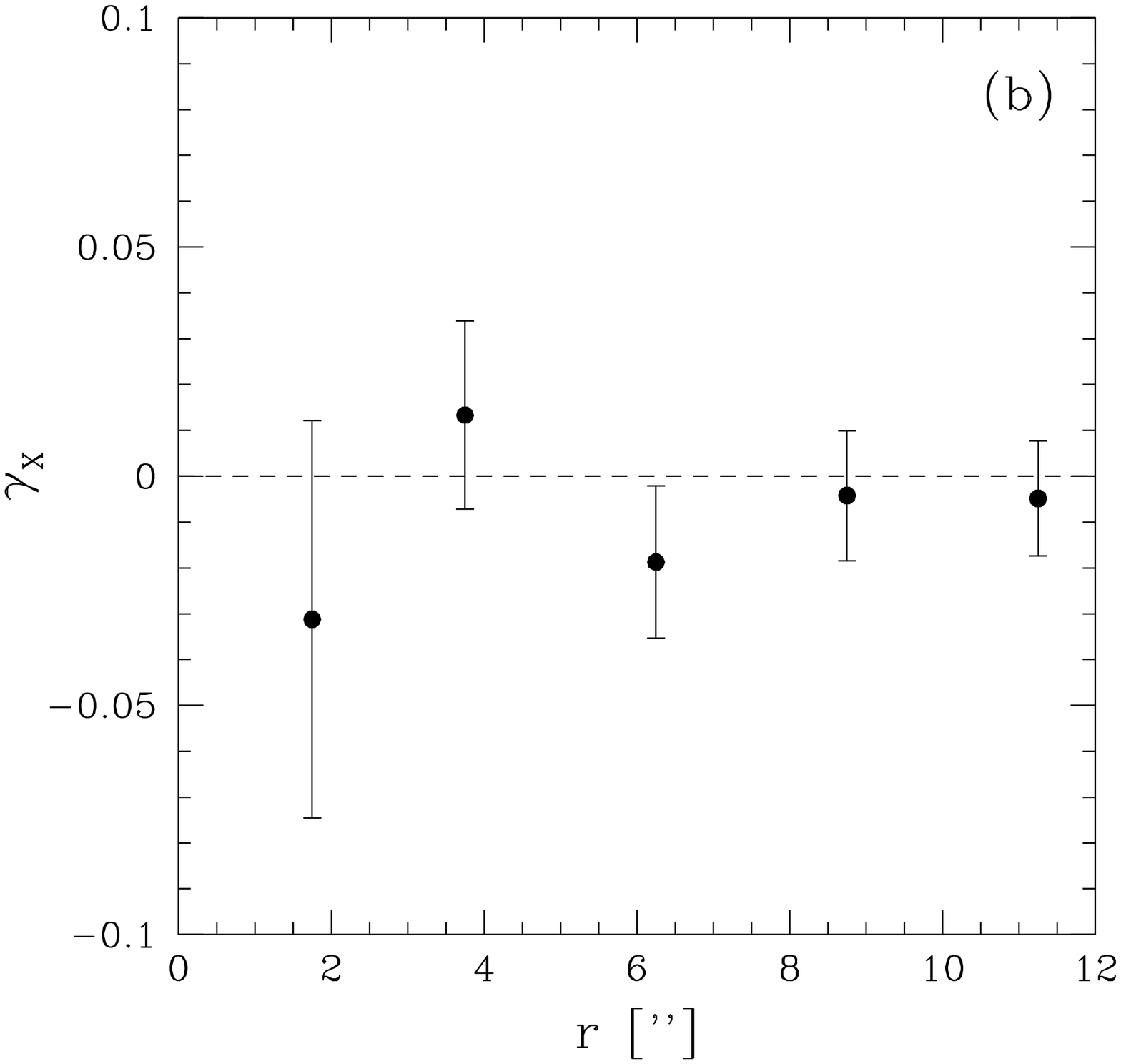}}
\begin{small}
\figcaption{(a) Average tangential shear around an ensemble of cluster 
galaxies. The signal induced by the smooth cluster mass distribution is 
removed from the observed signal, using the mass reconstruction. The signal 
for each lens galaxy has been scaled to match the signal for a $L_*$ 
($L_{*B}=8\times 10^{10} h_{50}^{-2} L_{B\odot}$) galaxy, assuming that mass 
scales with $\sqrt L_B$. The solid line is a singular isothermal sphere 
fitted to the measured points. It corresponds to a velocity dispersion of 
$\sigma=203\pm33$ km/s. (b) Average tangential shear around the cluster 
galaxies when the phase of the shear is increased by $\pi/2$.
\label{galgal}}
\end{small}
\end{center}}

Converting the lensing signal into a mass estimate requires knowledge of
the redshift distribution of the faint sources. We have explored the use
of photometric redshift distributions inferred from the Northern 
(Fern{\'a}ndez-Soto et al. 1999) and Southern (Chen et al. 1998) Hubble Deep 
Fields. Comparison of the results for the two fields suggest field to field
variations which lead to a 10\% systematic uncertainty in the mass estimate 
for MS~1054-03. Our results for the mass and mass-to-light ratio of MS~1054-03
agree well with other estimators, suggesting that these photometric redshift 
distributions are a fair approximation of the true distribution. This gives 
us confidence that we are able to measure a well calibrated weak lensing mass 
for this high redshift cluster.

Using the $\zeta_c$ statistic (Clowe et al. 1998), we find a mass
$(1.2\pm0.2)\times10^{15} h_{50}^{-1}{\rm M}_\odot$ within an aperture of 
1~$h_{50}^{-1}$ Mpc radius. Under the assumption of an isothermal mass 
distribution, this corresponds to a velocity dispersion of $1311^{+83}_{-89}$ 
km/s. A fit of an isothermal sphere model to the observed tangential 
distortion at large radii, yields a velocity dispersion of $1365\pm84$ km/s. 
However, this result depends on the assumed radial surface density profile.
The uncertainties in the mass estimates only reflect the contribution from 
the intrinsic ellipticities of the sources. The weak lensing mass estimate 
is in good agreement with the X-ray estimate from Donahue et al. (1998) 
and the observed velocity dispersion (Tran et al. 1999; van Dokkum 1999).

The observed average mass-to-light ratio in the passband corrected $B$
band is $124\pm17 h_{50} {\rm M}_\odot/{\rm L}_{B\odot}$ (measured 
within a 1~$h_{50}^{-1}$~Mpc radius aperture). We use the results from van 
Dokkum et al. (1998) to correct our estimate for luminosity evolution. The 
resulting mass-to-light ratio of 
$269\pm37~h_{50} {\rm M}_\odot/{\rm L}_{B\odot}$ can be compared to results
obtained for nearby clusters, and is in excellent agreement with the findings
of Carlberg et al. (1997).

Our high resolution mass reconstruction shows three distinct peaks in the 
mass distribution, which are in good agreement with the irregular light 
distribution. The results of LK97 lacked the resolution to detect this 
substructure. In projection the clumps are approximately 500 $h_{50}^{-1}$ 
kpc apart, and we estimate the timescale for virialization to be at on 
the order of $\sim$ 1 Gyr. Thus the cluster should be virialized at a 
redshift $z \sim 0.65$. We found that the clumps are of roughly equal mass, 
with corresponding velocity dispersions of $\sim 700$ km/s.

To study the masses of the cluster members, we have measured the average 
tangential shear around a sample of 241 bright cluster galaxies. Assuming the 
Faber-Jackson scaling relation we find a velocity dispersion of $203\pm33$ 
km/s for an $L_*$ galaxy (${\rm L}_{*B}=8\times10^{10}h_{50}^{-2}
{\rm L}_{B\odot}$ for MS~1054-03). The mass-to-light ratio within a radius of 
$20 h_{50}^{-1}$ kpc is ${\rm M}/{\rm L}_B=7.5\pm2.4 h_{50}
{\rm M}/{\rm L}_{B\odot}$. Correcting this result for luminosity evolution
yields ${\rm M}/{\rm L}_B=17\pm5 h_{50}{\rm M}/{\rm L}_{B\odot}$.

The number of known massive high redshift clusters of galaxies is increasing 
rapidly. The results presented here demonstrate that these systems can be 
studied with sufficient signal-to-noise to be of interest for studies of 
cluster formation. The improved sampling and throughput of the Advanced Camera
for Surveys, that will be installed on the HST in the near future, will be 
very useful for weak lensing analyses of high redshift clusters of galaxies, 
especially when the results are combined with forthcoming X-ray observations. 
The combination of detailed weak lensing analyses and X-ray observations by 
Chandra and XMM will provide important information on the thermodynamics of 
cluster formation.

\acknowledgments

The authors thank Pieter van Dokkum for making the data available in reduced
form, which has been invaluable for this analysis.

\clearpage
\appendix

\section{Error analysis}

The weak lensing signal is obtained by averaging the shape measurements of
a number of faint galaxies. The noise in the shape estimate differs from
object to object, and therefore it is important to weight the measurements 
properly.

For bright galaxies the error on the shear measurement is dominated by
the intrinsic ellipticity distribution of the galaxies. At fainter magnitudes, 
shot-noise increases the uncertainty in the measurements. As these galaxies
require large seeing corrections, the measurement error on the distortion 
can be quite high. This is illustrated in figure~\ref{scatter}. 

Figure~\ref{scatter}a shows the observed scatter in the polarization as a
function of apparent magnitude in the $F814W$ band. As fainter objects
suffer more from the circularization by the PSF one would expect
the scatter to decrease towards fainter magnitudes. However, the 
increasing contribution by shot noise (indicated by the solid line)
cancels out the effect of the circularization.
Figure~\ref{scatter}b shows the observed scatter in the ellipticities (shear)
of the objects. The scatter at faint magnitudes increases rapidly due to the 
combination of shot noise and large PSF corrections. The solid line indicates
the expected scatter due to the combination of shot noise and PSF correction.

\noindent We estimate the weighted mean of the distortion using:

\begin{equation}
\langle g_i \rangle = \frac{\sum w_n e_{i,n}/P^\gamma_n}{\sum w_n},
\end{equation}
\noindent where the weight $w_n$ is the inverse of the variance of the
distortion:
\begin{equation}
w_n=\frac{1}{\sigma_g^2}=\frac{P_\gamma^2}{\langle \gamma_0^2 \rangle 
P_\gamma^2+\langle \Delta e^2\rangle},
\end{equation}
\noindent where $\langle \gamma_0^2 \rangle$ is the scatter due to the 
intrinsic ellipticity of the galaxies, $P_\gamma$ is the pre-seeing shear 
polarizability and $\langle \Delta e^2 \rangle^{1/2}$ is the error estimate 
for the polarization, which is derived below.

This procedure, however, gives only the optimal estimate for the average 
ellipticity, but does not account for the fact that the nearby bright galaxies
are lensed less efficiently than faint high redshift galaxies. To get the 
best lensing estimate it is therefore also necessary to weight with $\beta$ 
as well. This is demonstrated in figure~\ref{weight}a, where we show the 
weight function as a function of apparent magnitude (solid line). The
dashed line shows the weight function hen we include the strength of the 
lensing signal. This effectively decreases the weight of bright (nearby) 
galaxies and increases the relative weight of the faint (distant) 
galaxies.

It is also interesting to see which magnitude range contributes most to
the measurement of the distortion. In figure~\ref{weight}b we plot the weight 
function for each magnitude bin times the number of galaxies in that bin. 
This gives the total contribution of each bin to the measurement of the 
distortion. Although the bright galaxies have the highest weight function (cf. 
figure~\ref{weight}a), they are not so numerous to contribute significantly to 
the lensing signal. At the faint end both incompleteness and increasing noise 
causes the contribution to decrease rapidly. Figure~\ref{weight}b also shows
that including the strength of the lensing signal $(\beta)$ changes the 
profile only slightly. It shows that galaxies with $F814W$ magnitudes between 
24 and 26.5 contribute most to the lensing signal.

\subsection{Estimating the measurement error on the polarization}

Here we derive an expression to estimate the error on the polarization 
measurement, using the observations.

The observed quadrupole moments are given by:
\begin{equation}
I_{ij}^{\rm obs}=\int d^2x x_i x_j W(\vec x) f_{\rm obs}(\vec x),
\end{equation}

\noindent which are combined to form the two-component polarization.
\begin{equation}
e_1 = \frac{I_{11}-I_{22}}{I_{11}+I_{22}}~{\rm and}
~e_2 = \frac{2I_{12}}{I_{11}+I_{22}}
\end{equation}

\noindent Due to noise, the true image $f_{\rm true}(\vec x)$ is changed to
the observed image $f_{\rm obs}(\vec x)$:

\begin{equation}
f_{\rm obs}(\vec x)= f_{\rm true}(\vec x) + \sigma b(\vec x) + d(\vec x)\sqrt{f(\vec x)},
\end{equation}

\noindent where we take $\langle b(\vec x)^2\rangle^{\frac{1}{2}}=
\langle d(\vec x)^2\rangle^{\frac{1}{2}}=1$. The first term in equation A5 
gives the true image, the second term is the noise in the background, and the 
third contribution comes from the shot noise due to the image. We therefore 
observe quadrupole moments given by:

\begin{equation}
I^{\rm obs}_{ij}=I^{\rm true}_{ij} + 
\int d^2x x_i x_j W(x) \left(\sigma b(\vec x) + d(\vec x)\sqrt{f(\vec x)}\right)
\end{equation}

\noindent As $\langle b(\vec x)\rangle=\langle d(\vec x)\rangle=0$, this shows
that $\langle I^{\rm obs}_{ij}\rangle=I^{\rm true}_{ij}$. Thus noise does not
introduce a bias in the average measurement of the quadrupole moments. However,
it does introduce a neglegible bias in the polarization.
We now can estimate the variance in the observed
\twocolumn

\vbox{
\begin{center}
\leavevmode
\hbox{%
\epsfxsize=\hsize
\epsffile[20 144 340 710]{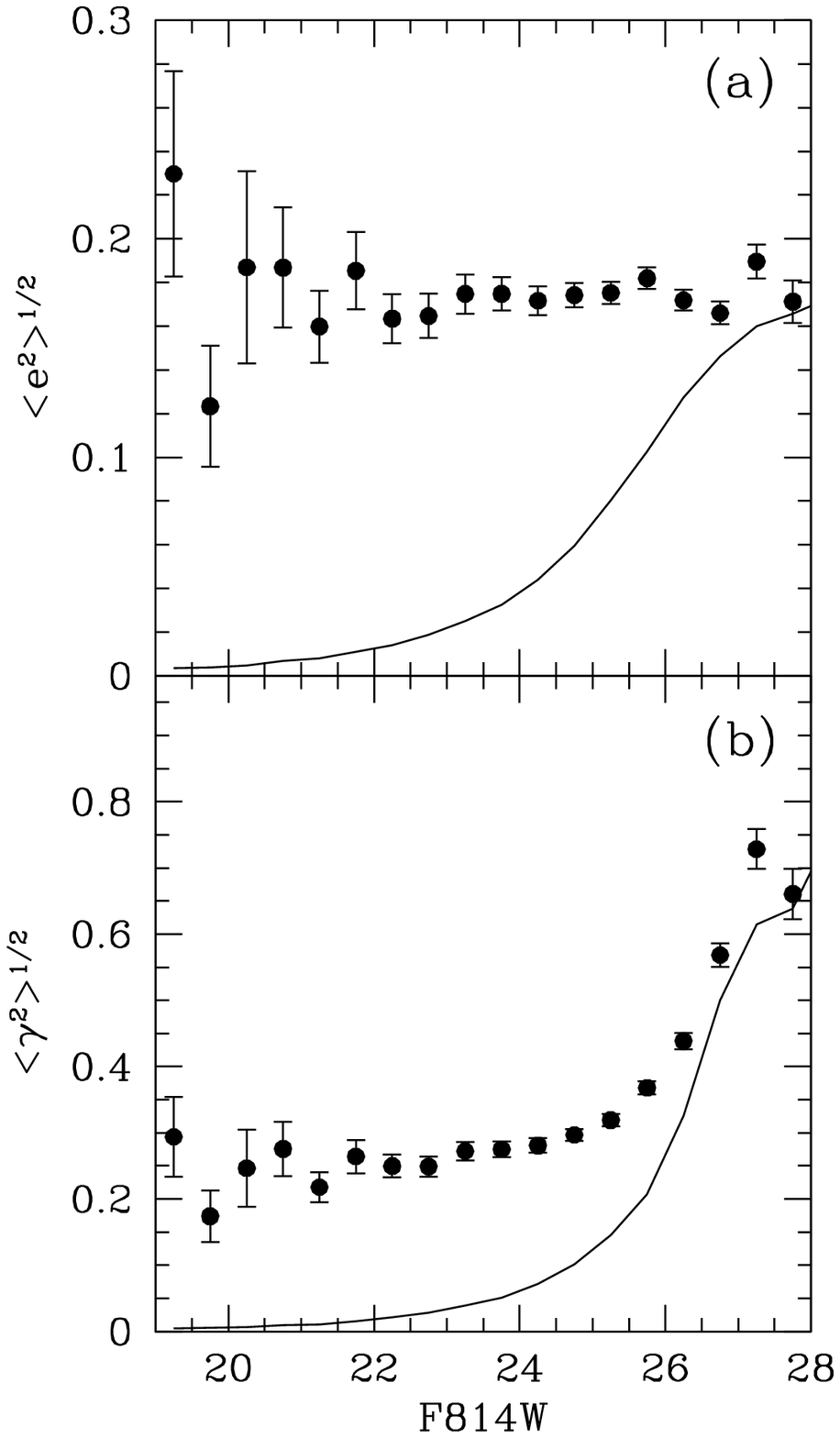}}
\begin{small}
\figcaption{(a) Plot of the observed scatter in the polarization as a function 
of apparent magnitude in the $F814W$ band. The scatter is fairly constant with 
apparent magnitude. The PSF circularizes the images of the faint galaxies, 
but the increasing scatter due to shot noise cancels out this effect. (b) 
Plot of the observed scatter in the ellipticities (shear) of the galaxies.
At bright magnitudes the scatter is dominated by the intrinsic ellipticities
of the galaxies. At fainter magnitudes the scatter increases rapidly due
to increasing shot noise and large PSF corrections. The contribution of
the shot noise to the scatter is indicated by the solid line.
\label{scatter}}
\end{small}
\end{center}}

\noindent quadrupole moments using
equation~A6 and noting that 
$\langle b(\vec x) b(\vec y)\rangle=\langle b(\vec x) b(\vec y)\delta(\vec x - \vec y)\rangle$,
$\langle d(\vec x) d(\vec y)\rangle=\langle d(\vec x) d(\vec y)\delta(\vec x - \vec y)\rangle$,
and $\langle b(\vec x) d(\vec y)\rangle=0$, where $\delta(\vec x - \vec y)$ is the 
Dirac delta function. This yields:

\begin{equation}
\langle \Delta I_{ij} \Delta I_{kl}\rangle=
\int d^2x x_i x_j x_k x_l W^2(\vec x)\left( \sigma^2+f(\vec x)\right)
\end{equation}

\vbox{
\begin{center}
\leavevmode
\hbox{%
\epsfxsize=\hsize
\epsffile[30 170 350 700]{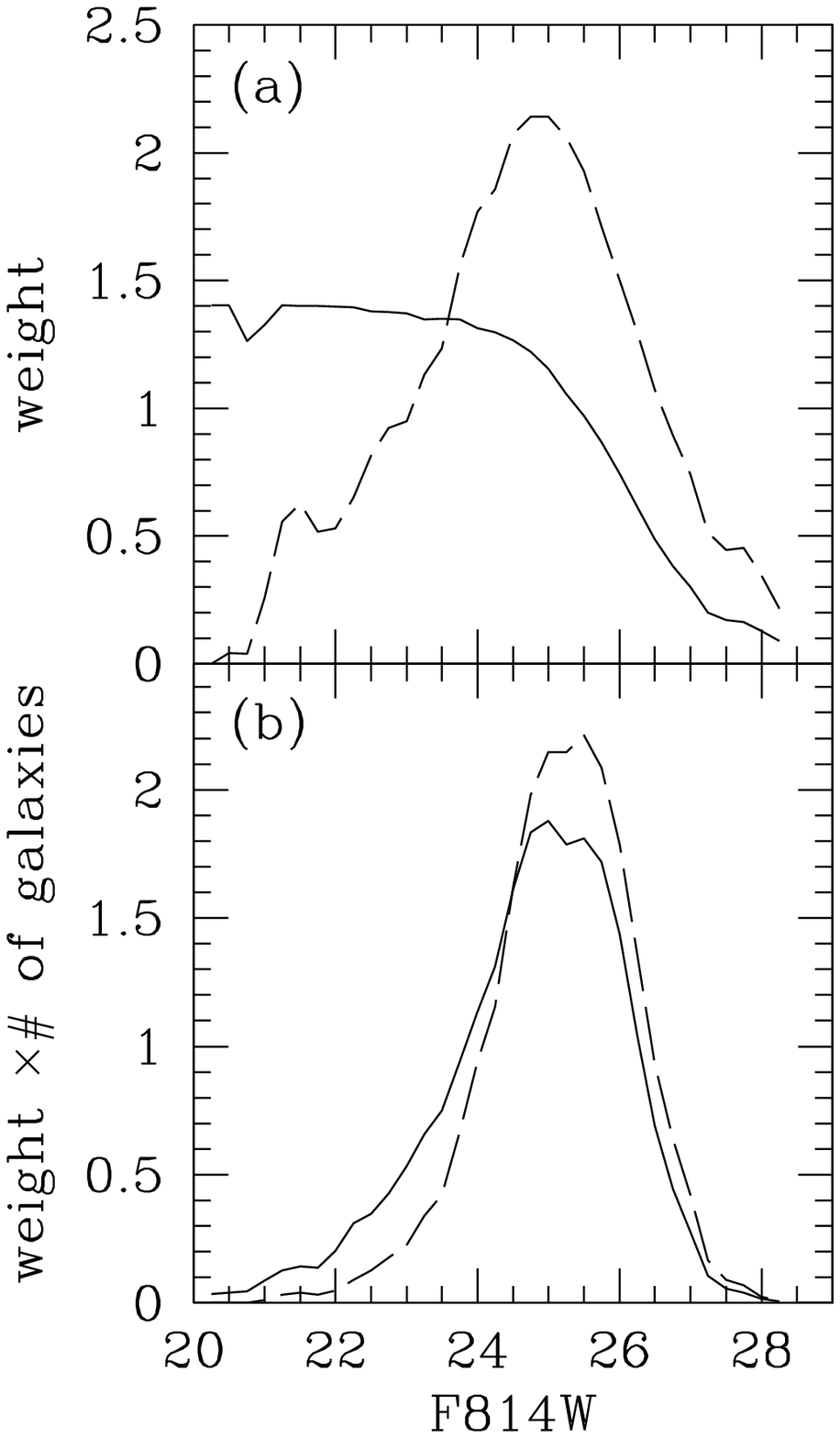}}
\begin{small}
\figcaption{(a) Weight function as a function of apparent $F814W$ 
magnitude (solid line). For bright galaxies the weight function is dominated 
by the intrinsic ellipticities of the galaxies. At fainter magnitudes, 
the weight function decreases rapidly due to the increasing measurement
error on the distortion. Including the strength of the lensing signal 
($\beta$) yields a different weight function (dashed line), which lowers 
the weight of bright (nearby) galaxies and thus increases the relative weight 
of faint (distant) galaxies. (b) Plot of the weight function times the number 
of objects in each magnitude bin as a function of apparent $F814W$ magnitude 
(solid line). This plot shows which magnitude bin contributes most to the 
signal. Although bright galaxies have the highest weight (cf. 
figure~\ref{weight}), they are not so numerous that they contribute 
significantly to the lensing signal. Including the strength of 
the lensing signal $(\beta)$ changes the profile mainly at the bright end, but
the differences are small. The lensing signal is dominated by galaxies with 
$F814W$ magnitudes between 24 and 26.5.
\label{weight}}
\end{small}
\end{center}}

We now want to estimate the errors on the measurements of the polarizations
$e_1$ and $e_2$ and express them in terms of the errors on the
quadrupole moments. This finally gives:

\begin{figure*}
\begin{center}
\leavevmode
\hbox{%
\epsfxsize=8cm
\epsffile{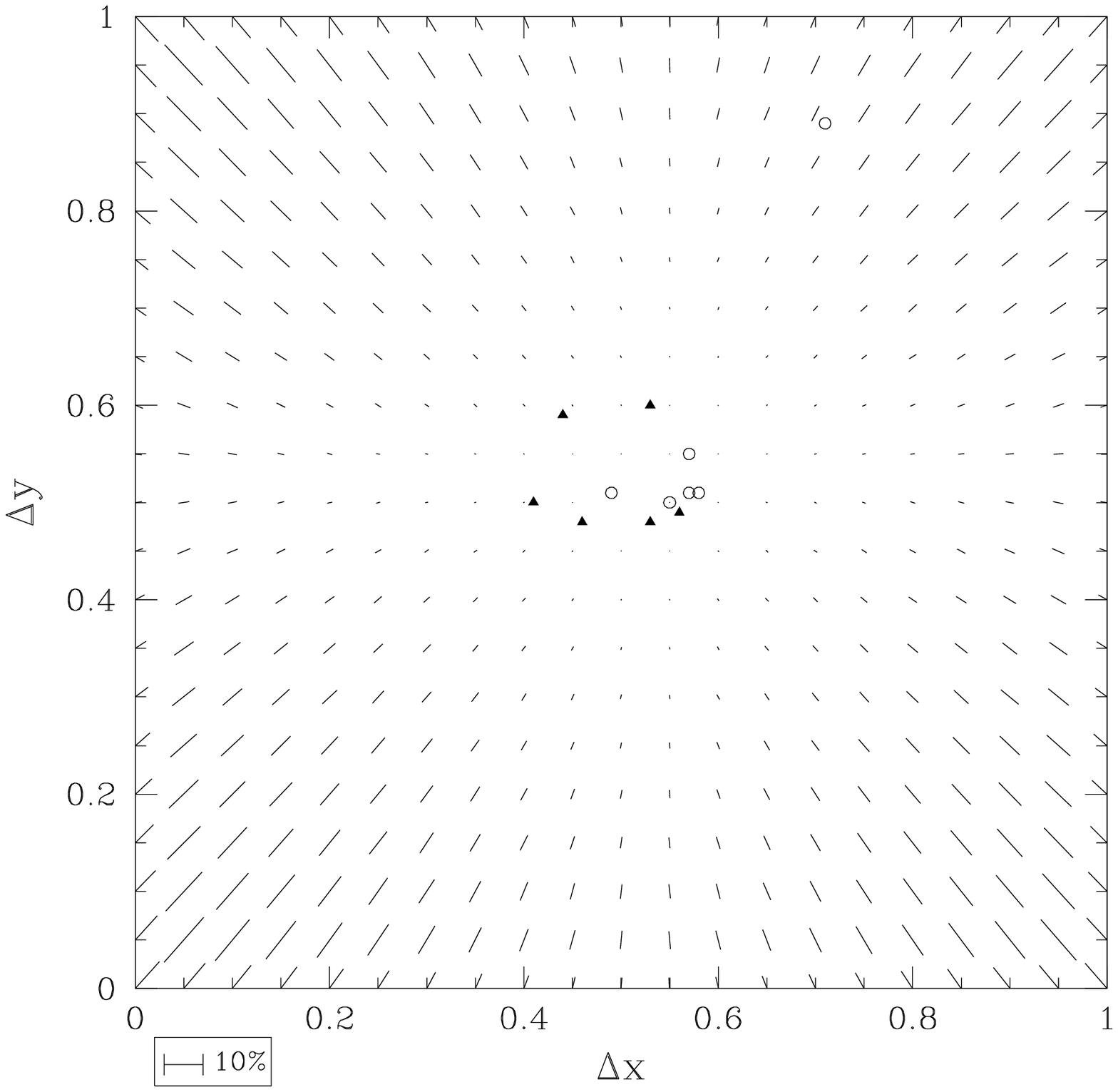}
\epsfxsize=8cm
\epsffile{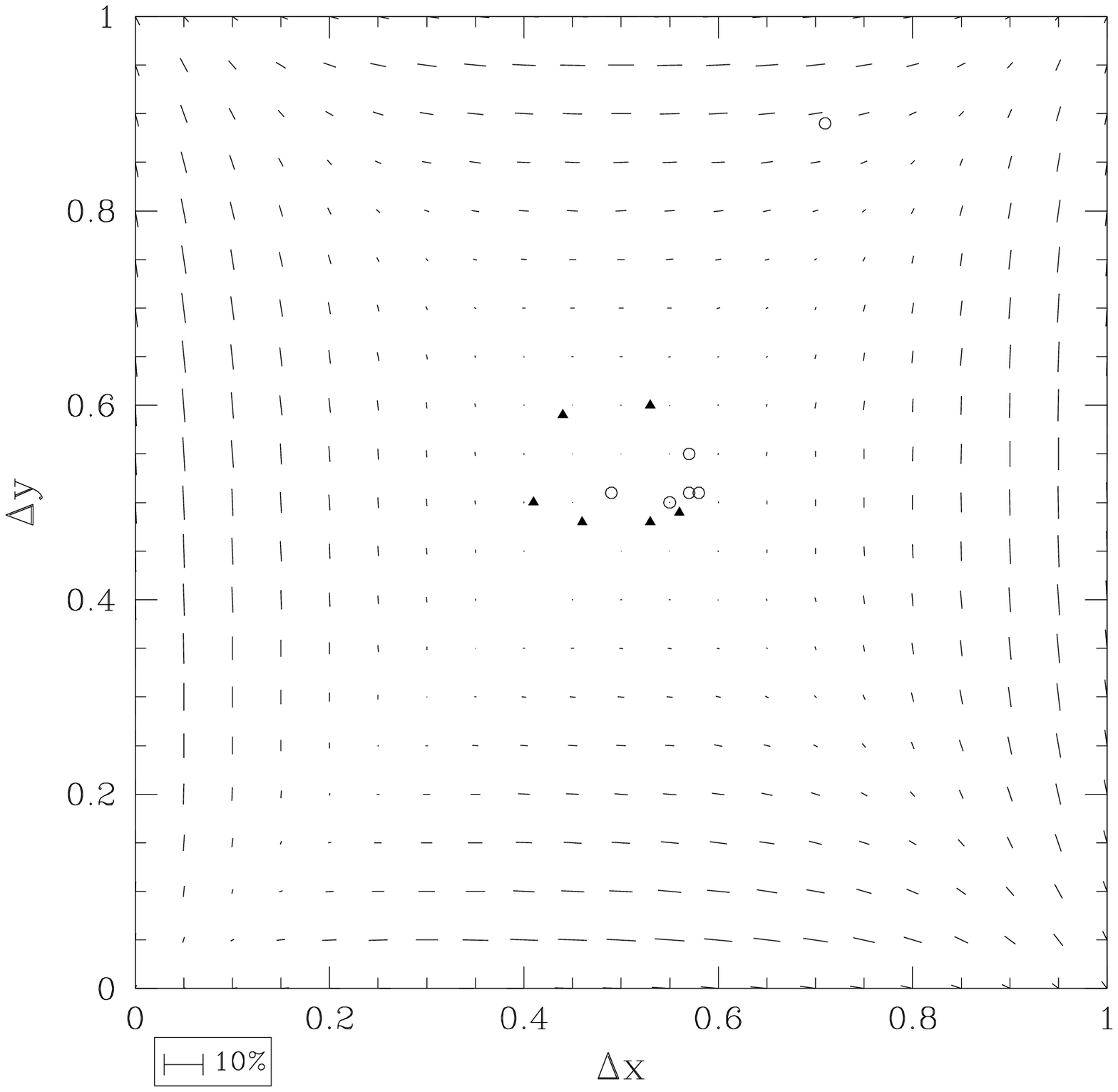}}
\begin{small}
\figcaption{Left panel: Ellipticity introduced by constructing an interlaced
images from images with imperfect offsets. The reference star (a Tiny Tim PSF)
is centered exactly on a pixel. This image is shifted $(\Delta x, \Delta y)$ 
and an interlaced image is created, assuming the shift was (0.5, 0.5) pixel; 
Right panel: The same procedure was repeated for the case in which the 
reference star is placed at the intersection of four pixels. The sticks 
indicate the direction and the size of the introduced ellipticity. If the 
shifts are half pixel, the measured ellipticity is correct. The offsets of our 
observations for $F814W$ (open circles) and $F606W$ (solid triangles) are also
indicated. For one $F814W$ pointing the offset is such that results of the 
interlaced images are biased.
\label{interlace}}
\end{small}
\end{center}  
\end{figure*}

\begin{eqnarray}
\langle \Delta e_1^2 \rangle & = & \frac{1}{(I_{11}+I_{22})^2}
[(1-e_1)^2 \langle \Delta I_{11}^2 \rangle+ (1+e_1)^2 \langle \Delta I_{22}^2\rangle\nonumber\\
& & \qquad -2(1-e_1^2)\langle \Delta I_{12}^2\rangle]
\end{eqnarray}

\noindent and

\begin{eqnarray}
\langle \Delta e_2^2 \rangle & = & \frac{1}{(I_{11}+I_{22})^2}
[e_2^2 (\langle \Delta I_{11}^2 \rangle+\langle \Delta I_{22}^2\rangle)\nonumber\\
& & \qquad + (4+2 e_2^2)\langle \Delta I_{12}^2\rangle \nonumber\\
& & \qquad -  4 e_2 (\langle \Delta I_{11}\Delta I_{12}\rangle 
+ \langle \Delta I_{22}\Delta I_{12})\rangle].
\end{eqnarray}

From simulations and real images we find that the errors on both
$e_1$ and $e_2$ are similar. This estimate for the error on the
polarization can be calculated from the observations and
allows us to properly weight the signals of the galaxies using
equation~A2.

\section{The effects of interlacing}

The images of small, faint galaxies in WFPC2 observations are
poorly sampled. HFKS98 showed that as a result the smallest objects cannot be
corrected reliably for the PSF effects.

However, several techniques can be used to improve the sampling of WFPC2 
observations by dithering the exposures (Fruchter \& Hook 1998). When the 
shifts are half pixel in both $x$ and $y$ (modulo integer pixel shifts) one 
can construct an interlaced image (Fruchter \& Hook 1998) which has a 
$\sqrt{2}$ better sampling.

Interlaced images were constructed from the observations of MS~1054-03.  
Doing so, we assumed that the offsets were exactly half pixel. Due to 
pointing errors and the camera distortion the real offsets are 
slightly different.

\begin{figure*}
\begin{center}
\leavevmode
\hbox{%
\epsfxsize=8cm
\epsffile{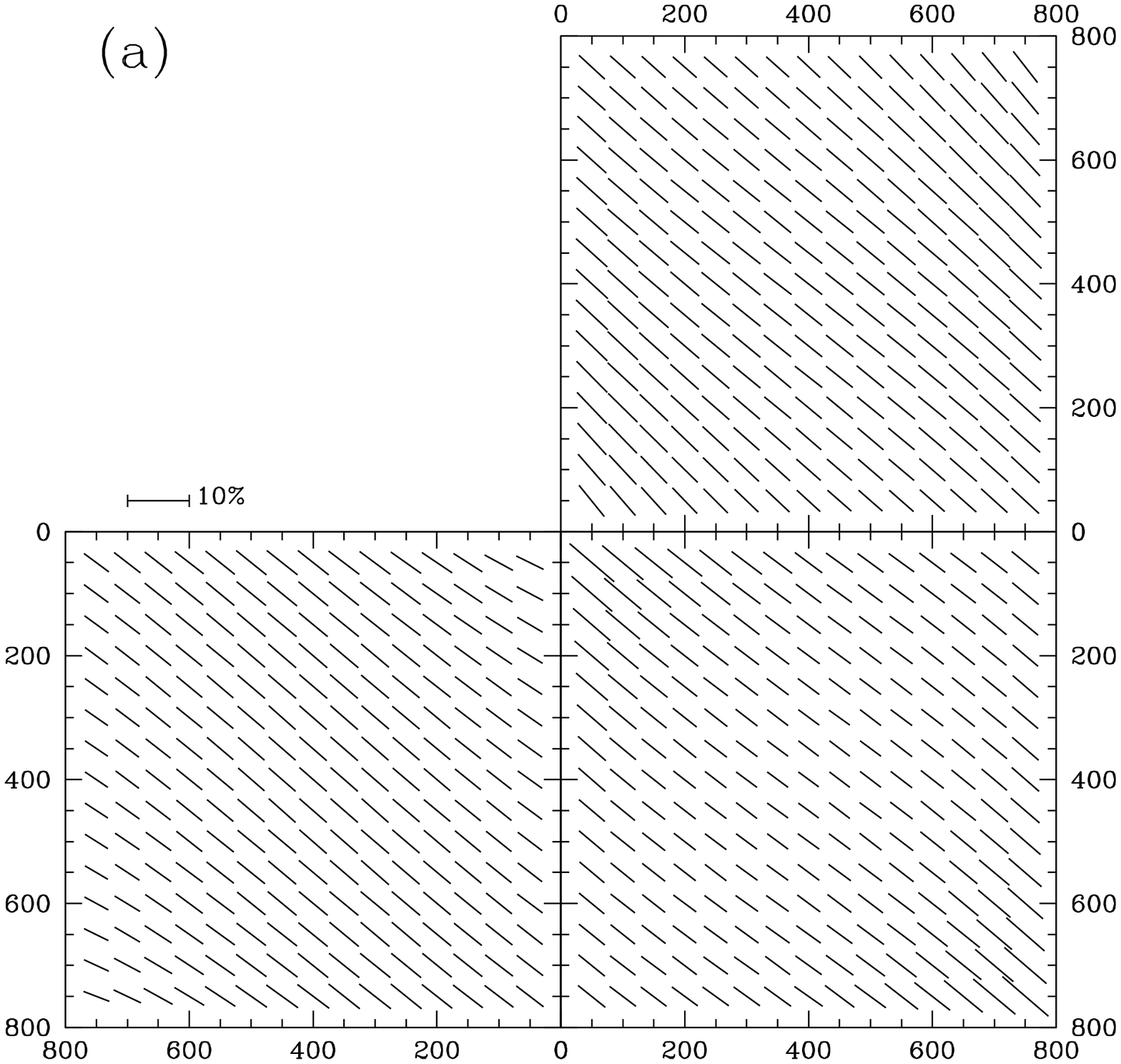}
\epsfxsize=8cm
\epsffile{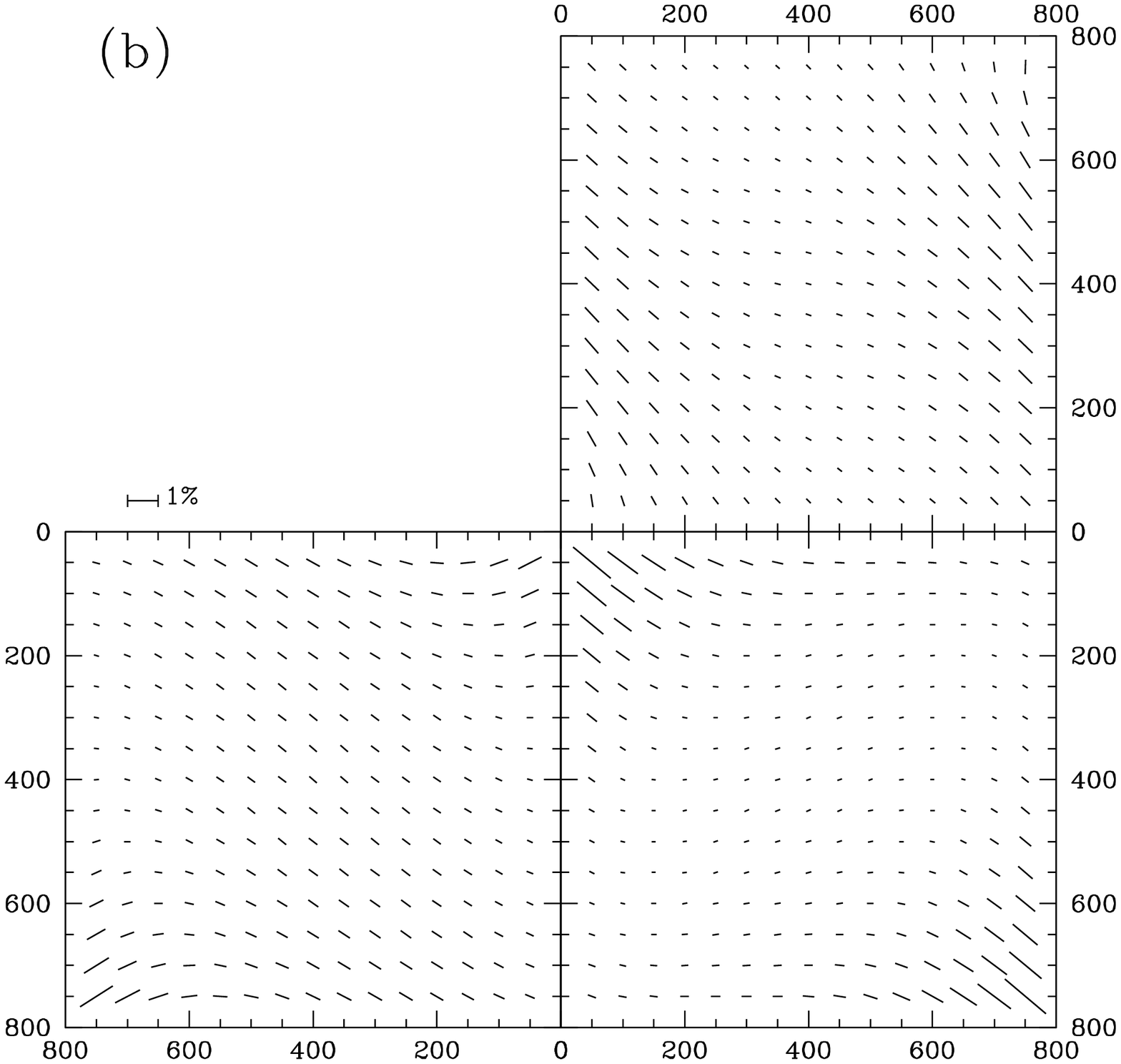}}
\begin{small}
\figcaption{Ellipticity introduced by imperfect offsets and WFPC2 camera 
distortion. (a): The introduced ellipticities for the worst offset in 
the $F814W$ filter $(\Delta x, \Delta y)=(5.71,5.89)$ as a function of 
position. In this case it is clear that interlacing introduces a bias in the 
observed shapes, and therefore these data (the $F814W$ image of the first
pointing) are not used in the weak lensing analysis. (b): The introduced 
ellipticity for the second worst offset (5.53, 5.60). In this case the bias 
is very small. Due to the telescope distortion the effect is somewhat larger 
at the edges of the chips. The sticks indicate the direction and the size of 
the introduced ellipticity.
\label{interlace2}}
\end{small}
\end{center}  
\end{figure*}

In this section we examine the effect of imperfect offsets on the shape 
measurements. To do so, we examine the change in the shape of a model
PSF. Although interlacing is not a convolution, it affects large
objects (as these are well sampled even before interlacing) 
less than small objects. By examining the effects on the PSF, we study 
a worst case scenario. The effects should be less for most
of the galaxies used in the weak lensing analysis.

We use a ten times oversampled model PSF calculated using Tiny Tim. 
This PSF is shifted and rebinned to WFPC2 sampling and convolved
with the pixel scattering function, in order to realistically mimic
the WFPC2 PSF. 

We start with a reference image in which the star is centered exactly
on a pixel. This image is shifted $(\Delta x, \Delta y)$ and an interlaced
image is created assuming the shift was exactly (0.5, 0.5) pixels.
The resulting image is analysed and the measured ellipticities are compared 
to the correct value. 

The change in ellipticity as a function of the applied offset is presented in 
figure~\ref{interlace} (left panel). When the reference star is placed at the
intersection of four pixels the results are different, as is indicated in
figure~\ref{interlace} (right panel). These two extreme cases should cover the 
full range of possible configurations.

In figure~\ref{interlace} we also plot the observed offsets for our data. The
open circles represent the offsets of the $F814W$ images, and the solid 
triangles show the offsets for the $F606W$ images. Only one $F814W$ pointing
has such a deviating offset that the shape measurements cannot be used for
our weak lensing analysis. In all other images the offsets are such that
no significant bias is introduced by the interlacing.

Even if the offset is perfect in the centre of the image, telescope distortions
will result in imperfect offsets at the edges.  We simulated the amplitude of 
the effect using the coefficients for the telescope distortions from Holtzman 
et al. (1995).

Figure~\ref{interlace2} shows the introduced ellipticity for our worst 
offset (5.71, 5.89). The bias in the observed ellipticities is clear in this 
case. These data (the $F814W$ image of the first pointing) were not used in 
the weak lensing analysis. The second worse offset (5.53, 5.60) is also shown 
in this figure. The ellipticity introduced in this case is small. Only at the 
edges of the chips the camera distortion introduces a small effect. As the 
effect is even smaller for galaxies, we feel certain that we can use the 
interlaced images for our weak lensing analysis.

\end{document}